\documentclass[a4paper]{article}
\usepackage[utf8]{inputenc}
\usepackage{graphicx}
\usepackage{tikz}
\usepackage{xargs}
\usepackage{ifthen}
\usepackage{amssymb}
\usepackage{xparse}
\usepackage{hyperref}
\usetikzlibrary{patterns}
\usetikzlibrary{math}
\usetikzlibrary{arrows.meta}
\usetikzlibrary{calc}
\usetikzlibrary{shapes.multipart}
\usetikzlibrary{shadows}
\usetikzlibrary{shadings}
\usepackage{cancel}
\usepackage{xstring}
\usepackage{xcolor}
\usepackage{soul}
\usepackage[font=footnotesize,width=13cm]{caption}

\hypersetup{
    colorlinks=true,
    linkcolor=black,
    filecolor=black,
    urlcolor=black,
    citecolor=black,
    bookmarks=true,
    bookmarksopen=true,
    pdftitle={MOON},
}

\tikzset{
diagonal fill/.style 2 args={fill=#2, path picture={
\fill[#1, sharp corners] (path picture bounding box.south west) -|
                         (path picture bounding box.north east) -- cycle;}},
reversed diagonal fill/.style 2 args={fill=#2, path picture={
\fill[#1, sharp corners] (path picture bounding box.north west) |- 
                         (path picture bounding box.south east) -- cycle;}}
}

\newcommand{\ifthen}[2]
{
\ifthenelse{#1}{#2}{}
}

\newcommandx*{\grille}[3][3=0,usedefault]
{
\ifthenelse{\equal{#3}{0}}
{\draw[white,fill=white,opacity=0] (-#1,-#2) rectangle (#1, #2);}
{\draw[step=0.2,draw=black!20,very thin] (-#1,-#2) grid (#1,#2);
\draw[step=1.,draw=black!60,very thin] (-#1,-#2) grid (#1,#2);
\draw[thick] (0,-#2) -- (0,#2);
\draw[thick] (-#1,0) -- (#1,0);}
}

\newcommand{\examColor}{green}
\newcommand{\treatColor}{red!80!white}
\newcommand{\infeColor}{teal}
\newcommand{\programColor}{blue!80!white}

\newcommand{\elColor}{gray}
\newcommand{\dataColor}{magenta}

\newcommand{\GetCategoryColor}[1]
{
\def\nodeColor{\elColor}
\ifthen{\equal{#1}{EXAM}}{\def\nodeColor{\examColor}}
\ifthen{\equal{#1}{INFERENCE}}{\def\nodeColor{\infeColor}}
\ifthen{\equal{#1}{TREATMENT}}{\def\nodeColor{\treatColor}}
\ifthen{\equal{#1}{PROGRAM}}{\def\nodeColor{\programColor}}
\ifthen{\equal{#1}{TREXAM}}{\def\nodeColor{white!40!black}}
}

\newcommandx*{\DataSpecSingle}[5]
{
\node (#3) at (#1, #2) [transform shape, rounded corners=0.5mm,rectangle, black, 
text centered, text width=6cm, scale=0.8, draw=\dataColor, fill=white!90!\dataColor]
{Dat\_Typ\_Id : {\color {\dataColor} #4} \par Raw Data Type : {\color {red} #5}};
}

\newcommandx*{\ResultSpec}[7][6=FromClass,7=1,usedefault]
{
\GetCategoryColor{#4}
\def\nodeName{#3}
\ifthenelse{\equal{#6}{FromClass}}
{}{\def\nodeName{#6}}
\node (\nodeName) at (#1, #2) 
[transform shape, rounded corners=2mm,rectangle, black, text centered, text width=5cm, scale=0.8, draw=\nodeColor, opacity=#7, fill=white!90!\nodeColor] 
{{\bf \StrSubstitute{#3}{~}{\_}}\par Category : {\color{\nodeColor!80!black} #4} \par\vskip 2pt Data : {\color{red} #5}};
}

\newcommandx*{\PrescRes}[8][7=No\_Param,8=1,usedefault]
{
\GetCategoryColor{#5}
\node (#3) at (#1, #2) [transform shape, rectangle, text centered, text width=4cm, scale=0.8, draw=\nodeColor, fill=white!90!\nodeColor] 
{{\bf \StrSubstitute{#3}{~}{\_}}\par\vskip 2pt Parameter : {\color{red} #7}};
\ResultSpec{#1 + 5.5}{#2}{#4}{#5}{#6}[][#8];
\draw [->, >=latex] (#3) -- (#4);
}

\newcommandx*{\PrescForRes}[7][6=No\_Param,7=1,usedefault]
{
\GetCategoryColor{#5}
\path (#4) ++ (#1, #2) node (#3)
[transform shape, rectangle, text centered, text width=4cm, scale=0.8, draw=\nodeColor, fill=white!90!\nodeColor] 
{{\bf \StrSubstitute{#3}{~}{\_}}\par\vskip 2pt Parameter : {\color{red} #6}};
\draw [->, >=latex] (#3.east) -- (#4);
}

\newcommandx*{\MakeDepText}[4][2=1,3=All,4=0,usedefault]
{
\def\DepText{#2 $\leq$ Nb (#1)}
\ifthen{\NOT \equal{#3}{All}}
 {
 \ifthenelse{\equal{#2}{#3}}
  {\def\DepText{Nb (#1) = #2}}
  {\def\DepText{#2 $\leq$ Nb (#1) $\leq$ #3}}
 }
\ifthen{\NOT \equal{#4}{0}}
 {
 \ifthenelse{\equal{#2}{#3}}
  {\def\DepText{Nb (#1) = #2 \par in the last #4}}
  {\def\DepText{#2 $\leq$ Nb (#1) $\leq$ #3 \par in the last #4}}
 }
}

\newcommandx*{\MakeDepAttributes}[4][2=1,3=All,4=0,usedefault]
{
\def\DepText{$\begin{array}{lcl}
             \mathrm{\bf Dep\_Typ\_Id} & : & \mathrm{#1.Type\_Id}\\
             \mathrm{\bf Card\_Min}    & : & \mathrm{#2}\\
             \mathrm{\bf Card\_Max}    & : & \mathrm{#3}\\
             \mathrm{\bf Validity}     & : & \mathrm{#4}
             \end{array}$}
}

\newcommandx*{\AddDep}[7][6=orange,7=0,usedefault]
{
\def\txtDepColor{black}
\ifthen{\equal{#7}{1}}{\def\txtDepColor{gray}}
\path (#3) ++  (#1, #2) node (Dep #4)
[transform shape, rectangle, \txtDepColor, text centered, text width=8cm, scale=0.75, draw=#6, fill=white!90!#6]
{\small{\bf #5}};
\draw [-{Ellipse[length=6pt,width=2pt]}] (Dep #4.east) -- (#3);
}

\newcommand{\opPreDep}{0.65}

\newcounter{ResNumber}
\newcounter{PreNumber}
\newcounter{nbPrescribed}
\newcounter{firstPreNum}

\def\PresentTime{0}
\def\TimeLineY{0}
\def\TimeLineX{12}
\def\ValMarkSize{0.45}
\def\ValMarkCenterShift{0.195}
\def\ProdMarkSize{0.225}
\def\ProdMarkCenterShift{0.0975}

\def\EHRMode{AllTimes}

\newcommand{\InitPreResUIDs}
{
\setcounter{ResNumber}{0}
\setcounter{PreNumber}{0}
}

\newcommandx*{\TimeLine}[2][1=12,2=0,usedefault]
{
\InitPreResUIDs
\def\TimeLineX{#1}
\def\TimeLineY{#2}
\setcounter{nbPrescribed}{0}
\ifthen{\NOT \equal{\EHRMode}{NoTimes}}
 {
 \draw [->, >=latex, very thick] (-#1, #2) -- node [midway](TimeLine){} (#1, #2);
 \node at (#1 + 0.2, #2 - 0.1)[transform shape, scale=0.8, anchor=north east]{\it Time};
 }
}

\newcommandx*{\PresentLine}[3][2=0.5,3=4.75,usedefault]
{
\def\PresentTime{#1}
\ifthen{\NOT \equal{\EHRMode}{NoTimes}}
{
 \draw [red!80!blue, thick, dashed] (#1,\TimeLineY) -- ++ (90:#3) node [scale=0.7,anchor=east,transform shape]{Present};
 \draw [red!80!blue,fill=red!80!blue] (#1,\TimeLineY+0.5) circle (1mm);
 \draw [red!80!blue,->, >=latex] (#1, \TimeLineY+#2) -- (-\TimeLineX, \TimeLineY+#2) node [transform shape,above=8pt,scale=0.7,right=10pt,anchor=west] {Past};  
 \draw [red!80!blue,->, >=latex] (#1, \TimeLineY+#2) -- (\TimeLineX, \TimeLineY+#2) node [transform  shape,above=8pt,scale=0.7,left=15pt,anchor=east] {Future}; 
 }
}

\newcommand{\ComputeResultCoordinates}[4]
{
\coordinate (ReDeb#1) at (#2, \TimeLineY);
\path (ReDeb#1) ++ (0:#3) coordinate (ReEnd#1);
\path (ReEnd#1) ++ (0:#4) coordinate (ReDel#1);
\path (ReDeb#1) ++ (0:#3/2.0) coordinate (ReVal#1);
\path (ReDeb#1) ++ (0:#3/2.0) coordinate (ReProd#1);
}

\newcommandx*{\ComputeInferenceValueCoordinates}[4][4=0,usedefault]
{
\path (ReVal#2) ++ (0,#3) coordinate (ReVal#1);
\ifthen{\NOT \equal{#4}{0}}
 {
 \coordinate (ReValFromDep#1) at (ReVal#1);
 \path (ReValFromDep#1) ++ (0:#4) coordinate (ReVal#1);
 }
}

\newcommand{\DrawLinkToMarker}[2]
{
\ifthen{\NOT \equal{\EHRMode}{NoTimes}}
 {
 \draw [\nodeColor,densely dotted,opacity=\globOp,] (Re#1.south) |- ($(Re#1.south) ! 0.15 ! (#2)$) -| (#2);
 }
}

\newcommandx*{\DrawValueSymbol}[3][3=0,usedefault]
{
\draw [#2,fill=white] (#1) ++ (0,-\ValMarkCenterShift) ++ (0,#3) --++ (60:\ValMarkSize) --++ (180:\ValMarkSize) -- cycle;
}

\newcommandx*{\DrawProdSymbol}[3][3=0,usedefault]
{
\draw [#2,fill=#2] (#1) ++ (0,-\ProdMarkCenterShift) ++ (0,#3) --++ (-60:\ProdMarkSize) --++ (180:\ProdMarkSize) -- cycle;
}

\newcommandx*{\DrawDeliverySymbol}[2]
{
 \draw[#2,fill=#2] (ReDel#1) circle (1.5mm);
 \path (ReDel#1) ++ (90:0.15) coordinate (MarkDel);
 \DrawLinkToMarker{Re#1}{MarkDel};
}

\newcommandx*{\DrawInferenceValue}[3][2=0,3=1,usedefault]
{
\ifthen{\equal{\EHRMode}{Values} \OR \equal{\EHRMode}{AllTimes}}
 {
 \def\lStyle{dashed}
 \ifthen{\equal{#3}{0}}{\def\lStyle{solid}}
 \node (ReValNode) at (ReVal#1) [transform shape,draw=\infeColor,fill=\infeColor!20!white,scale=0.6,rounded corners=2pt , \lStyle]{\texttt{Res #1} Value};
 \path (ReValNode.south) ++ (270:3mm) coordinate (ReValCenter#1);
 \draw[\infeColor,thick,densely dotted] (Re#1.south) |- (ReValCenter#1); 
 \DrawValueSymbol{ReValCenter#1}{\infeColor}
 \ifthen{\NOT \equal{#2}{0}}
  {
  \node at (ReValFromDep#1) [transform shape,draw=white,fill=white,scale=0.6,rounded  corners=2pt ,\lStyle]{\color{white}\texttt{Res #1} Value};
  \node (ReValNodeDep) at (ReValFromDep#1) [transform shape,draw=\infeColor,fill=\infeColor!20!white,scale=0.6,rounded  corners=2pt ,\lStyle,opacity=0.25]{Res #1 Value};
  \path (ReValNodeDep.south) ++ (270:3mm) coordinate (ReValFromDep#1);
  \draw[\infeColor,densely dotted,thick,opacity=0.25] (Re#1.south) |- (ReValFromDep#1); 
  \DrawValueSymbol{ReValFromDep#1}{white}
  \DrawValueSymbol{ReValFromDep#1}{\infeColor}
  \path (ReValNodeDep.north) coordinate (From);
  \path (ReValNode.north) coordinate (To);
  \path ($(From)!0.5!(To)$) ++ (90:0.2) coordinate (MidProj);
  \draw[->,>=latex,gray] (From) |- (MidProj) -| (To);
  }
 }
}

\newcommand{\DrawServiceMarkers}[1]
{
\ifthen{\equal{\EHRMode}{Services} \OR \equal{\EHRMode}{AllTimes} \OR \equal{\EHRMode}{ValDel}}
 {
 \ifthen{\NOT \equal{\EHRMode}{ValDel}}
  {
  \draw [\nodeColor,very thick] (ReDeb#1) ++ (270:0.3) coordinate(MarkStart) --++(90:0.6);
  \DrawLinkToMarker{#1}{MarkStart};
  \draw [\nodeColor,thick] (ReEnd#1) ++ (-0.2,-0.3) -- ++(0:0.2) coordinate (MarkEnd) -- ++(90:0.6) -- ++(180:0.2);
  \DrawLinkToMarker{#1}{MarkEnd};
  }
 \draw[\nodeColor,fill=\nodeColor] (ReDel#1) circle (1.5mm);
 \path (ReDel#1) ++ (90:0.15) coordinate (MarkDel);
 \DrawLinkToMarker{#1}{MarkDel};
 }
}

\newcommand{\DrawValueMarker}[1]
{
\ifthen{\equal{\EHRMode}{Values} \OR \equal{\EHRMode}{AllTimes} \OR \equal{\EHRMode}{ValDel} \OR \equal{\EHRMode}{ValProd}}
 {
 \draw [\nodeColor,densely dotted,thick,opacity=0.5*\globOp] (Re#1.south) -- (ReVal#1);
 \DrawValueSymbol{ReVal#1}{\nodeColor}[\ValMarkCenterShift]
 }
}

\newcommand{\DrawProdMarker}[1]
{
\ifthen{\equal{\EHRMode}{Productions} \OR \equal{\EHRMode}{AllTimes} \OR \equal{\EHRMode}{ValProd}}
 {
 \draw [\nodeColor,densely dotted,opacity=0.5*\globOp] (Re#1.south) -- (ReVal#1);
 \DrawProdSymbol{ReVal#1}{\nodeColor}[\ProdMarkCenterShift]
 }
}
\newcommandx*{\SmallRes}[9][8=0,9=0,usedefault]
{
\GetCategoryColor{#6}
\def\lStyle{dashed}
\ifthenelse{\equal{#8}{0}}{\def\lStyle{solid}}{}
\ifthenelse{\equal{#8}{2}}{\def\globOp{0.25}}{\def\globOp{1.0}}
\begin{scope}[opacity=\globOp]
\stepcounter{ResNumber}
\ComputeResultCoordinates{\theResNumber}{#2}{#3}{#4};
\ifthen{\NOT \equal{#1}{0}}
 {
 \path (ReDeb\theResNumber) ++ (0,-0.3) coordinate (ReStart);
 \draw [gray,densely dotted,->,>=latex] (Pr#1.east) -| (ReStart);
 }
 {
 \path (ReVal\theResNumber) ++ (90:#5) node (Re\theResNumber) [transform shape,scale=0.6,rectangle,draw=\nodeColor, fill=\nodeColor!20!white,text width=2cm, text centered, rounded corners=2pt,\lStyle]{\color{black} \texttt{Res \arabic{ResNumber}}  \par {\scriptsize #7}};
 }
\DrawProdMarker{\theResNumber}
\DrawServiceMarkers{\theResNumber}
\ifthen{\NOT \equal{#6}{INFERENCE}}{\DrawValueMarker{\theResNumber}}
\let\lStyle\undefined
\ifthen{\equal{#8}{2}}
 {
 \draw[red,thick] (Re\theResNumber.north east) -- (Re\theResNumber.south west);
 \draw[red,thick] (Re\theResNumber.north west) -- (Re\theResNumber.south east);
 \draw[white,very thick,opacity=1-\globOp] (ReDeb\theResNumber) ++ (270:0.3) --++(90:0.6);
 }
\end{scope}
}

\newcommandx*{\SmallPresc}[7][3=0,4=None,5=None,6=0.6,7=0.65,usedefault]
{
\GetCategoryColor{#1}
\stepcounter{PreNumber}
\stepcounter{nbPrescribed}
\ifthenelse{\equal{#3}{0}}
{\path (\PresentTime, \TimeLineY) ++ (270:\thenbPrescribed*1.35-0.0) coordinate (PrescLoc);}
{\path (\PresentTime, \TimeLineY) ++ (270:#3) coordinate (PrescLoc);}
\ifthenelse{\equal{#5}{None}}
{\def\prescText{\color{black}\texttt{Presc~\arabic{PreNumber}}\nodepart{two}\footnotesize #2 \\ \color{red}#4}}
{\def\prescText{\color{black}\texttt{Presc~\arabic{PreNumber}}\nodepart{two}\footnotesize #2 \\ \color{red}#4\\\color{black}#5}}
\ifthenelse{\equal{#1}{TREXAM}}
 {
 \node (Pr\arabic{PreNumber}) at (PrescLoc) [transform shape,opacity=1,scale=#7,rectangle split, rectangle split parts=4,rectangle split ignore empty parts, rectangle split draw splits=false,draw=\nodeColor, diagonal fill={\treatColor!20!white}{\examColor!20!white},text width=2cm, text centered,inner sep=1.75mm] {\prescText};
 }
 {
 \node (Pr\arabic{PreNumber}) at (PrescLoc) [transform shape,opacity=1,scale=#7,rectangle split, rectangle split parts=4,rectangle split ignore empty parts, rectangle split draw splits=false,draw=\nodeColor, fill=\nodeColor!20!white,text width=2cm, text centered,inner sep=1.75mm] {\prescText};
 }
\draw[\nodeColor,densely dashed] (Pr\arabic{PreNumber}.text split west) -- (Pr\arabic{PreNumber}.text split east);
}

\newcommandx*{\PrescAtDate}[3][3=Current,usedefault]
{
\ifthenelse{\equal{#3}{Current}}
{}{\GetCategoryColor{#3}}
\draw [white, fill=white, opacity=0.0] (#2, \TimeLineY) coordinate (PrDate) {} circle (0.1mm);
\draw [very thick,densely dotted,\nodeColor] (PrDate) ++ (270:0.3) coordinate (MarkStart) -- ++(90:0.6);
\draw [gray,densely dotted,->,>=latex] (Pr#1.east) -| (MarkStart);
}

\newcommandx*{\Validity}[6][5=1,6=0.75,usedefault]
{
\def\vColor{cyan!90!black};
\def\vSymb{\large ?};
\ifthenelse{\equal{#5}{1}}
 {\def\vColor{green!70!black};
  \def\vSymb{\large \checkmark}}
 {\ifthenelse{\equal{#5}{2}}
  {\def\vColor{red!70!black};
   \def\vSymb{\large $\times$}}
  {}
 }
\path (ReDeb#1) ++ (270:#2) coordinate (ValStart);
\path (ValStart) ++ (180:#3) coordinate (ValEnd);
\draw [\vColor,densely dotted] (ValStart) -- (ReDeb#1);
\draw [\vColor,densely dotted] (ValEnd) -- ++(90:#2);
\draw [\vColor, dashed, |-|] (ValStart) -- node [fill=white, transform shape, above,scale=#6]{\color{\vColor}\bf #4 \vSymb} (ValEnd);
}

\newcommand{\depTo}{Re}

\newcommandx*{\SmallDep}[5][3=0,4=1,5=0,usedefault]
{
\ifthenelse{\equal{#4}{1}}
{\def\lStyle{dashed}}
{\def\lStyle{solid}}
\ifthen{\equal{#5}{0}}
{\draw[-{Kite[length=6pt,width=2pt]},opacity=\opPreDep](Re#1) -- (\depTo#2);
 \ifthenelse{\equal{#3}{0}}
 {}
 {\ifthen{\equal{\EHRMode}{Values} \OR \equal{\EHRMode}{AllTimes}}
  {
  \path (ReVal#1) ++ (90:#3) coordinate (ReValCoord);
  \node (ReVal#2) at (ReValCoord) [transform shape,draw=\infeColor,fill=\infeColor!20!white,scale=0.6,rounded corners=2pt,\lStyle]{\texttt{Res #2} Value};
  \path (ReVal#2.south) ++ (270:3mm) coordinate (ReValCenter#2);
  \draw[\infeColor,densely dotted] (Re#2.south) |- (ReValCenter#2); 
  \DrawValueSymbol{ReValCenter#2}{\infeColor}
  }
 }
}
\ifthen{\equal{#5}{1}}
 {\draw[-{Kite[length=6pt,width=2pt]},opacity=\opPreDep*0.5, dashed](Re#1) -- (\depTo#2);}
\ifthen{\equal{#5}{2}}
 {\draw[-{Kite[length=6pt,width=2pt]},opacity=\opPreDep, dashed,brown](Re#1) -- (\depTo#2);}
}

\newcommandx*{\SmallProgram}[5][2=None,3=0.6,4=0,5=0.8,usedefault]
{
\setcounter{nbPrescribed}{0}
\setcounter{firstPreNum}{0}
\foreach \Pre/\PreParam/\PreCat/\PreDates/\ServID in #1
 {
 \SmallPresc{\PreCat}{\Pre}[0][\PreParam][\ServID][#3][#5]
 \ifthen{\equal{\thefirstPreNum}{0}}{\setcounter{firstPreNum}{\thePreNumber}}
 \ifthen{\equal{#2}{None}}
  {\foreach \pDate in \PreDates {\PrescAtDate{\thePreNumber}{\pDate}[\PreCat]}}
 }
\ifthenelse{\equal{#2}{None}}
{
}
{\foreach \pNum/\pCat/\pDate in #2 {\PrescAtDate{\pNum}{\pDate}[\pCat]}}
\path (Pr\thefirstPreNum.north west) ++ (-3pt,3pt) coordinate (Haut);
\path (Pr\thePreNumber.south east) ++ (3pt,-3pt) coordinate (Bas);
\draw[\programColor, thick, rounded corners=2pt] (Haut) rectangle (Bas);
\draw[\programColor,thick] (Pr\thefirstPreNum.north) ++(90:3pt) -- (\PresentTime,\TimeLineY - 0.3) --++(0,0.6);
}

\sethlcolor{yellow!50!white}

\def\MoonArticleMode{Draft}

\begin{document}
\title{MOON, an Operational Modeling of the Medical Pathway}
\author{Jean Hommet, and Daniel Cussol}
\date{LPC CAEN, Normandie Univ, ENSICAEN, UNICAEN, CNRS/IN2P3, 6 boulevard
Maréchal Juin, Caen, 14050 France\\ ~ \\ \today}

\maketitle

\abstract{
The use of digitized medical records is growing in importance.
Unfortunately digitized medical records suffer from a lack of coordination
between medical actors and medical departments, furthermore the proposed
standards focus more on data than on medical practices themselves. In this
context, reconstructing a patient's global medical history is virtually
impossible. Even worse, the management of this global medical pathway falls to
the patient, who lacks the necessary medical knowledge to handle it
efficiently. We propose a new framework named MOON whose underlying concepts
and structures allow for a specification of medical practices across virtually
any medical field and then can describe any medical pathway involving the
specified practices. This approach also enables the creation of comprehensive,
patient-centered medical records that can be shared among all medical
stakeholders. Adopting this type of framework would help improve the efficiency
of care and, consequently, patient health.

}


\tableofcontents

\newpage

\section{Introduction}
\label{introduction}

This article presents a conceptual medical framework called MOON (\emph{Medical
Operational ONtology}), designed to describe a patient's medical pathway and to
structure and organize its medical record.

This article is the first of a serie which will present successively :

\begin{itemize}
\item
   the low level concepts of the framework used to structure the overall
   patient pathway (this article),
\item
   a higher level concept, Protocol, which can describe generically
   any medical scenario,
\item
   Ada\_Moon, a simulation oriented implementation of MOON in Ada language,
\item
   the ``Worm model'', an illustrative application example within Ada\_Moon
   that simulates virtual patients undergoing radiotherapy, proton therapy,
   and carbon-ion therapy.
\end{itemize}

A MOON file is a patient-centered medical record, the use of which could lead
to major advancements at all level of medicine:

\begin{itemize}
\item
   Primary uses: care provided to each patient,
\item
   Secondary uses: statistical studies and numerical simulations.
\end{itemize}

Before presenting the MOON framework, we will provide an overview of the different
uses of the current medical records and their limitations.


\subsection{Current Medical Records}
\label{current-medical-records}

Patient records are an integral part of the software used by healthcare
professionals to manage their services. To be effective, this software is
designed to meet the specific needs of each type of actor or facility: general
practitioners, specialists, nurses, physiotherapists, hospitals, etc. Each
patient's pathway is described in detail from the professional's perspective,
i.e., according to the logic of managing its own service. These records
therefore perfectly meet the internal needs of the service.

Thus, any patient managed by multiple medical actors has its pathway fragmented
across different professional software systems. The corresponding records are
therefore partial and, due to their varied perspectives and formats, cannot be
combined to reconstruct \textbf{a global and cross-functional pathway}.

Heterogeneous systems can still interoperate thanks to exchange standards and
terminologies such as FHIR \cite{FHIRhome}, SNOMED-CT \cite{SNOMEDhome}, LOINC
\cite{LOINChome}, or ICD-10 \cite{ICD10home}, which allow data sharing and
facilitate collaboration between practitioners. Unfortunately, this type of
interoperability remains ad-hoc and only concerns atomic aspects of the
records, never the entire pathway.

The centralized patient record, implemented in some countries like France
\cite{MonEspaceSante} or Germany \cite{ElektronischePatientenakteEPA}, changes
little: the supported data formats are limited to PDF and common image formats
like JPEG. Professional data are not systematically added, and the patient can
fill them as they wish. Its organization and management thus fall to the
patient, making this record inoperable.

Despite the widespread use of computing among all actors, patients remain
responsible for tracking their appointments and informing new practitioners
about their medical histories. Nothing has changed for patients, who must still
collect their data and manage their own records.

In summary, current medical records come in two forms:

\begin{itemize}
\item
   Partial records held by healthcare professionals: digitized, optimized,
   localized, service-specific, and inaccessible outside the service,
\item
   A ``complete'' patient record: on paper or in PDF/JPEG format, managed by the patient.
\end{itemize}


\subsection{Limitations of Current Records}
\label{limitations-of-current-records}


\subsubsection{Secondary Uses of Records}
\label{secondary-uses-of-records}

Secondary use refers to the application of anonymized records outside their care
context. This most often involves statistical studies covering a large number
of records. These studies may concern any form of pathology or treatment, as
well as related aspects such as billing, financing, or resource allocation.
They help advance medical knowledge or evaluate medical or organizational
policies.

There are two types of studies:

\begin{enumerate}
\def\labelenumi{\arabic{enumi}.}
\item
   Prospective studies: conducted within a service while patients are being
   treated, they are use cases of the service. They therefore have precise
      data, but their statistical scopes are very limited.
\item
   Retrospective studies: carried out using archived records, they are not use
   cases of the services. They thus allow for broader statistics, but suffer
   from data quality issues.
\end{enumerate}

Here are some inherent obstacles to the retrospective use of records:

\begin{itemize}
\item
   The study is not an initial use case for the records,
\item
   Data may be missing,
\item
   A large part of the data is in the form of unstructured text,
\item
   The relationships between data are not explicit in most cases (structure, cause, effect, order, dependency, etc.),
\item
   Temporal aspects are often imprecise or absent,
\item
   Records come from heterogeneous sources (e.g., cardiology departments from different hospitals),
\item
   Each patient's record consists of multiple sources (e.g., general practitioner and cardiologist for the same patient).
\end{itemize}

The value of a retrospective study heavily depends on the quality and interpretability of the data. This is a well-documented problem.

International collaborations such as STROBE
\cite{elmStrengtheningReportingObservational2007} and RECORD
\cite{benchimolREportingStudiesConducted2015} provide recommendations for
conducting such studies. They emphasize the need to explicitly describe the
methods used to extract source data and manage missing data. In 2013, Hersh
\cite{hershCaveatsUseOperational2013} compiled a list of warnings about the use
of such data, showing how previous studies could have been mistaken. Since
then, the list of recommendations continues to grow through numerous works,
including these three examples:

\begin{itemize}
\item
   Kohane \cite{kohaneWhatEveryReader2021} illustrates the complexity of the task through many cases,
\item
   Kim \cite{kimChallengesOpportunitiesElectronic2024} analyzes statistical biases caused by errors,
\item
   Qian \cite{qianHowDeepYour2025} shows that recent deep learning-based methods
   for attributing missing values still heavily depend on preprocessing and a priori
   knowledge of the data.
\end{itemize}

Thus, outside their initial context, collecting and interpreting data often
proves difficult, if not impossible.

One of the most demanding secondary uses is simulating a patient's pathway from
its medical record. To accurately reproduce certain aspects of the pathway, a
precise and rigorous understanding of the data is required, which explains why
this practice is rarely developed.

Patient pathway simulation should at least be used internally. However, this
review of patient flow logistics methods \cite{zamaniOptimizingPatientFlow2025}
does not mention it, even though almost all examples in this review were
conducted inside a medical department.


\subsubsection{Patient Uses}
\label{patient-uses}

The electronic record benefits the internal management of services but not the
patient, who remains responsible for managing its own paper or PDF/JPG record.
This impractical and unreliable record is the only global version of
its pathway serving as a reference outside of services --- \textbf{The
patient's record supposedly for primary use has all the disadvantages
associated with retrospective uses}.

The patients must act as intermediaries between services: scheduling
appointments, managing their medical records, and presenting results and
reports to new practitioners. These tasks can be burdensome and complex,
leading to misunderstandings and errors. Indeed, most patients lack medical
expertise and are unaware of the importance of certain results in their medical
pathways.

The consequences for their health are inevitably negative, with direct effects
(stress, fatigue) and indirect effects (duplicate exams, delays, or
inappropriate care). As a result, these effects worsen with the patient's
health and vulnerability, and are maximized in emergency situations.


\subsection{A Unified Patient-Centered Record}
\label{a-unified-patient-centered-record}


\subsubsection{EHR Standards}
\label{ehr-standards}

These issues have long been known and will be resolved when the medical record
follows the patient across all services. Such a record is called an
\textbf{EHR} (Electronic Health Record), whereas the internal records of
services are called \textbf{EMR} (Electronic Medical Record)
\cite{ElectronicHealthRecords}. Kalra \cite{kalraElectronicHealthRecords1994}
already presented the advantages of EHRs in 1994 during the presentation of
European research and standardization projects on the subject. Among these was
GEHR (\emph{Good Electronic Health Record})
\cite{griffithPortableCommunicativeArchitecture1995}, launched in 1992, which
aimed to create a shared, patient-centered record. This work laid the
foundations for current EHR projects and standards, such as ISO-13606
\cite{ISO13606Standard}, openEHR \cite{bealeOntologybasedModelClinical2007},
and HL7-RIM \cite{ISOHL721731}. Their specifications have evolved over the
years through mutual inspiration and harmonization efforts. Together, they
represent a considerable amount of work, but these proposals remain marginal in
practice and only benefit locally.
\ifthenelse{\equal{\MoonArticleMode}{IEEEJBHI}}
{}
{\paragraph{}}
\label{impossible} 
The following lines can be read in the ISO-13606 standard:
\begin{quote}
   \emph{It is impossible to represent the highly
   complex, highly dynamic, multi-disciplinary/multi-domain healthcare system
   by one domain‘s terminology/ontology or - even worse - by using ICT
   ontologies.}
\end{quote}

This is one of the main reasons why all three solutions rely on a two-level
architecture. The first level consists of an abstract model (called the
reference or information model), associated with a set of generic tools
allowing each concept of the model to be derived in a controlled manner. The
second level includes many varied domain models built using the tools of the
first level by experts in different medical specialities.

The information models of HL7-RIM and ISO-13606 are technical data models
without medical specificity --- the ISO standard even states that its model can
accomodate any domain, citing particle physics as an example. The openEHR
reference model, on the other hand, would be a complete operational medical
model if many of its specifications were not optional.

The medical pathway is not taken into account by any of the three reference
models. In all three standards, the pathway can be considered in the definition
of the artifacts of the second modeling level, or at the application level via
interoperability interfaces. In any case, the resulting pathways are always
partial, specific, and unknown to the reference model.

As a result, the three proposals allow for abstract exploration of the record
(through the simple use of the reference model), but must rely directly on
specialized medical models to develop most functionalities. Ultimately, these
three reference models alone, and therefore without a consistent medical
pathway, do not allow for the implementation of truly operational use cases.


\subsubsection{MOON, a Reference Model: Medical and Operational}
\label{moon-a-reference-model-medical-and-operational}

Contrary to the claims of the ISO-13606 authors (\ref{impossible}), it is
possible to represent the healthcare system using a single model at the patient
level. This is what our conceptual MOON model proposes, representing the
patient's pathway through a very reduced set of abstractions.

For comparison, MOON retains the conceptual foundations of openEHR --- namely
\emph{the clinical investigation model} and \emph{the clinical information
model} \cite{bealeOntologybasedModelClinical2007} --- but in an extremely
compact modeling based on the \emph{acts} of HL7-RIM \cite{ISOHL721731}.

MOON's framework also consists of two levels. At the
first level, the \textbf{reference model} describes the \textbf{patient's
medical pathway} in an abstract way. The second level allows medical experts to
specify the \textbf{concrete elements of the pathway}. In doing so, the medical
community defines its own \textbf{medical model}.

As with the three models, the two-level architecture allows for control and
validation of data. Additionally, MOON makes it easier for medical experts to
specify content. Indeed:

\begin{itemize}
\item
   The number of concepts is very reduced,
\item
   The complexity of pathways is already managed by the reference model.
\end{itemize}

Finally, the abstract modeling of the pathway provides access to new functionalities:

\begin{itemize}
\item
   A wealth of simulation possibilities,
\item
   Internal organization and management of a service's workflows,
\item
   Publication of service offers,
\item
   Most secondary uses,
\item
   Systemic management of the record on behalf of the patient (data access, appointment scheduling, etc.).
\end{itemize}

The \textbf{MOON reference model} can be seen in two parts:

\begin{itemize}
\item
   The \textbf{base model}, mainly consisting of the following concepts:

   \begin{itemize}
   \item
      \emph{Prescription} and \emph{Result}, which are the elementary components of pathways,
   \item
      \emph{Dependency}, which defines dependency links between types of elements,
   \item
      \emph{Service}, which identifies an actor and defines its roles,
   \end{itemize}
\item
   \textbf{Model extensions} with the following concepts:

   \begin{itemize}
   \item
      \emph{Protocol}, which implements all kinds of scenarios,
   \item
      \emph{Unit}, which handles the internal functioning of a \emph{Service}.
   \end{itemize}
\end{itemize}

The base model allows for the description of any medical pathway; its
extensions make the model operational in practice.

This article presents the base model in detail; the extensions will be the
subject of future articles. Furthermore, this presentation focuses solely on
the concepts, without addressing other important aspects of shared records,
such as legislation, versioning, security, or distributed programming.

\newpage
\section{MOON, Base Model}
\label{moon-base-model}


\subsection{Fundamental Principles of the Medical Pathway}
\label{fundamental-principles-of-the-medical-pathway}

\emph{This section presents the principles of modeling the medical pathway in
broad terms. It introduces the categorization of pathway elements as well as
the relationships that connect them.}

The reference model is a model of the medical pathway. We define the medical
pathway as the \textbf{set of medical actions} performed over time, structured
by the \textbf{set of relationships} that exist between these actions.

These \textbf{actions} fall into four categories\footnote{The framework extension
provides a fifth category called \emph{Unit\_Side}, reserved for internal
service use, whose actions do not appear in the EHR but only in the service's
EMR. This type of use is not developed in this article.}:

\begin{itemize}
\item
   \textbf{\emph{Exam}}: observation made on the patient (temperature, image, etc.)
\item
   \textbf{\emph{Treatment}}: action performed on the patient (medication intake, surgery, radiotherapy, etc.)
\item
   \textbf{\emph{Inference}}: deduction made from previous results (diagnosis, organs contouring, etc.)
\item
   \textbf{\emph{Program}}: modification of the pathway (scheduling or unscheduling of actions)
\end{itemize}

The \textbf{relationships} between actions are of two kinds:

\begin{itemize}
\item
   \textbf{Scheduling relationships}, which show the origin of each pathway
   element and highlight decision chains,
\item
   \textbf{Dependency relationships}, which impose dependencies between action
   types and show which previous actions a given action depends on in the
   pathway (e.g., an organ contouring inference depends on an imaging exam
   on which the inference is based).
\end{itemize}

The four categories reflect the fundamental activities on which the logic of
medical pathways is based: \textbf{observe, treat, deduce, and decide}. Each
category has specificities that constrain the corresponding actions. Here are
two examples:

\begin{enumerate}
\def\labelenumi{\arabic{enumi}.}
\item
   Each action in the \emph{Exam} or \emph{Treatment} category must be performed
   in the patient's presence, whereas \emph{Inference} and \emph{Program} only
   require the pathway,
\item
   An \emph{Inference} action requires prior actions on which the inference is
   based.
\end{enumerate}

Thus, from these two statements, we can deduce some \textbf{operational rules}
of the pathway:

\begin{itemize}
\item
   It is not possible to schedule an exam or treatment if the patient is not
   available,
\item
   Scheduling an exam or treatment makes the patient unavailable during the
   exam or treatment,
\item
   Multiple inferences or scheduling actions can be performed at the same time,
   regardless of the patient's availability at that same time,
\item
   Scheduling an inference follows the identification of the actions on which
   it is based,
\item
   Every inference in the pathway points to the actions it depends on.
\end{itemize}

At this level of abstraction, many more rules can be stated. \textbf{The set of
rules constrains the pathway, gives it properties, and enables
functionalities.}

Dependencies and actions (classes) are specified at the second level of the
model by medical domain experts. Specifications are done by category, with the
constraints of each. In the pathway, dependency relationships between actions
(instances) form trees.

However, \emph{Program} actions cannot be specified at the second level by
experts. All actions in this category are already defined as \emph{system
actions} in the reference model. Nevertheless, including the \emph{Program}
category at the same conceptual level as the other categories makes the pathway
model uniform, continuous, and dynamic. Scheduling thus becomes recursive (with
programs of programs of programs, etc.). The pathway accounts for decisions
with scheduling relationships that also appear in a tree-like form.

In the end, the pathway presents the continuity of medical actions over time,
including \emph{Program} actions, which are treated in the same way as
\emph{Exam}, \emph{Treatment}, and \emph{Inference}. The specific rules of the
different categories apply to the specification of action classes and their
instances in the composition of the pathway. Thus, everything that is scheduled
is necessarily consistent, the pathway is always complete, and highlights the
logic of its progression through trees of scheduling and dependencies, without
leaving any a priori.


\subsection{Scheduling}
\label{scheduling}

\emph{This section introduces the concepts that represent actions, shows how
they are scheduled, and the resulting relationships.}

Each action is characterized by an \textbf{instruction} and an \textbf{expected
outcome}. Instructions are scheduled in the \textbf{future} and produce
expected outcomes at the scheduled times. The realization of an expected
outcome is done based on its instruction, at the \textbf{present}, when the
scheduled time is reached.

The entire pathway is fully described based on the \textbf{elementary
instruction/outcome pair} provided by the two abstract concepts
\textbf{\emph{Prescription}} and \textbf{\emph{Result}}, which respectively
embed a \textbf{\emph{parameter}} and a \textbf{\emph{data}}.

\emph{Results} are distributed over time and provide information about their
\emph{Prescription}. Thus, in the \textbf{future} as in the \textbf{past}, the
pathway accurately accounts for what has been scheduled. The \emph{data} of the
\emph{Result} is filled in at the \textbf{present} date according to the
instructions defined by the \emph{Prescription} and its \emph{parameter}. This
same data becomes accessible in the \textbf{past}.

The concepts \emph{Prescription} and \emph{Result} are the
\emph{Record\_Elements} that make up the pathways (and thus the records).

Any scheduling is itself ``prescribed'' using the following \emph{Program}
category pair:

\begin{itemize}
\item
   \textbf{\emph{Program\_Prescription}}, whose parameter is a
   \textbf{\emph{Protocol}},
\item
   \textbf{\emph{Program\_Result}}, whose data is a
   \textbf{\emph{Program\_Data}}.
\end{itemize}

\textbf{\emph{Protocol}} and \textbf{\emph{Program\_Data}} form the
\textbf{high-level instruction/outcome pair}:

\begin{itemize}
\item
   \textbf{\emph{Protocol}} is an abstract concept that allows scenarios to be
   created from \emph{Prescriptions} and \emph{Protocols} (recursive
   concept).
\item
   \textbf{\emph{Program\_Data}} is the result of the scheduling of the
   protocol; it is a simple structure grouping the \emph{Prescriptions} of
   the protocol and their scheduling dates.
\end{itemize}

The prescription of the \emph{Protocol} scenario produces a
\emph{Program\_Data} program, taking into account the patient's and service's
availability. The pathway is modified by adding the
\emph{Program\_Prescription} and its \emph{Program\_Result}, as well as all the
\emph{Prescriptions} and \emph{Results} of this scheduling.

A \emph{Protocol} allows for the creation of a wide variety of scenarios,
including protocols of protocols. Scenarios can include repetitions and
conditions associated with their execution. This is a complex concept whose
operation will be presented with the model extensions, and which can be omitted
to present the base model.
\ifthenelse{\equal{\MoonArticleMode}{IEEEJBHI}}
{}
{\paragraph{}}
\label{PP}
For the rest of this article, we will simply admit that protocols can be
nested, and we will limit usage examples to the \emph{Protocol\_Program}
protocol. This is the simplest of the \emph{Protocols}, as it simply involves
stating the \emph{Program\_Data} directly.

The scheduling relationships, presented in a parent/child form, allow for the
definition of trees of \emph{Record\_Element} instances:

\begin{itemize}
\item
   A \emph{Prescription} knows its children \emph{Results} (multiple possible),
\item
   A \emph{Result} knows its parent \emph{Prescription},
\item
   A \emph{Prescription} knows its parent \emph{Program\_Result},
\item
   A \emph{Program\_Prescription} knows its child \emph{Program\_Result}
   (only one),
\item
   The parent of a \emph{Program\_Prescription} is:

   \begin{itemize}
   \item
      itself, when this prescription is a prescription made directly by an
      actor on a patient (initial protocol),
   \item
      a \emph{Program\_Result}, when the protocol to be scheduled is contained
      within another.
   \end{itemize}
\end{itemize}

Each prescription by an actor on a patient adds new \emph{Record\_Elements} to
the pathway, whose scheduling links form a tree starting from this same
prescription. Thus, every element of the pathway allows access to the entire
chain of decisions that scheduled it. For example, for an exam result, one can
access the protocol of the initial prescription, all intermediate protocols
leading to it, and the exam's parameterization.


\subsection{Dependencies}
\label{dependencies}

\emph{This section presents dependency relationships, how they are stated, and
how they are translated in the pathway.}

When an action B depends on an action A, it means that action A occurs before
action B and that the result of B depends on that of A. For example, an organ
contouring inference B depends on an imaging exam A.

MOON places this type of relationship among the foundations of the pathway
model. Any activity that depends on another must be defined as such in order to
validate or invalidate its scheduling. Similarly, all preliminary results of a
given result must appear in the pathway and be explicitly identified. There
must be no a priori assumptions. Indicating that one result depends on another
is not an option but a requirement that is realized on two levels:

\begin{itemize}
\item
   \textbf{\emph{Dependency}}: expression of a dependency of a \textbf{type} of
   \emph{Prescription} with respect to a \textbf{type} of
   \emph{Record\_Element},
\item
   \textbf{\emph{Pre-Results}}: attribute of an \textbf{instance} of
   \emph{Result}, which lists the \textbf{instances} of preliminary
   \emph{Results} determined from a dependency of its \emph{Prescription}.
\end{itemize}

\emph{Dependency} is a simple structure, grouping:

\begin{itemize}
\item
   \emph{Dep\_Typ\_Id}: the identifier of the \emph{Record\_Element} type on
   which the dependency is based,
\item
   \emph{Card\_Min}: the minimum number of preliminary results,
\item
   \emph{Card\_Max}: the maximum number of preliminary results,
\item
  \emph{Validity}: the maximum age of each of the preliminary results with
   respect to the considered scheduling date (a deadline not to be
   exceeded).
\end{itemize}

A prescription can depend on multiple types of results. In other words, it can
have multiple dependencies. A list of \emph{Dependency} is thus defined for
each type of \emph{Prescription}.

A dependency can require multiple results; therefore, scheduling a given
\emph{Prescription} at a given date implies identifying a list of valid
preliminary \emph{Results} for each of its \emph{Dependency}.

For scheduling at date \emph{Dp}, a \emph{Pre\_Result} at date \emph{Dpr} is
valid with respect to a \emph{Dependency} when:

\begin{itemize}
\item
   the type of this \emph{Pre\_Result} is \emph{Dep\_Typ\_Id} when
   \emph{Dep\_Typ\_Id} identifies a \emph{Result}; otherwise, when
   \emph{Dep\_Typ\_Id} identifies a \emph{Prescription}, the type of this
   \emph{Pre\_Result} is the result type of that \emph{Prescription}
\item
   \emph{Dp - Dpr} $\leq$ \emph{Validity}.
\end{itemize}

A list of \emph{Pre\_Results} is valid with respect to a \emph{Dependency} if
it contains between \emph{Card\_Min} and \emph{Card\_Max} elements, each of them being valid.

Finally, scheduling a \emph{Prescription} at a given date is only possible if
all its dependencies are satisfied. The lists identified for this purpose
become the lists of \emph{Pre-Results} of the produced \emph{Result}.

In the end, scheduling each \emph{Result} requires identifying and validating
its \emph{Pre-Results} based on the constraints expressed once for all in the
specification of the \emph{Prescription}. The pathway is thus made consistent
and complete (without a priori). We know the premises of each \emph{Result} as
well as the roles it can play in the future.

Like a \emph{Prescription}, a \emph{Protocol} is also scheduled by declaring
the preliminary results it depends on. One might think that scheduling a
\emph{Protocol} comprising a dozen \emph{Prescriptions} requires significantly
more work. However, the internal dependencies of a \emph{Protocol} are fully
managed by the protocol itself, so that the number of dependencies of a
\emph{Protocol}, regardless of its size, is on average comparable to that of a
\emph{Prescription}. In the end, the vast majority of dependency relationships
inscribed in a pathway are realized automatically by the base model.


\subsection{Raw Data Packaging}
\label{raw-data-packaging}

\emph{In this section, we present the systematic packaging of raw data, the
purpose of which is to standardize and control access to them.}

Raw medical data is packaged in \emph{MOON\_Data}, each type of which is
identified by a \emph{Dat\_Typ\_Id}. \emph{MOON\_Data} is an introspective
abstraction whose internal structure is represented by paths accessing
\emph{MOON\_Data}. \emph{MOON\_Data} is divided into two families:
\emph{Single} and \emph{Composite}. New types are created using a class
template to which the necessary information is provided.

The template for a \emph{Single} requires:

\begin{itemize}
\item
   The identifier of the new type (\emph{Dat\_Typ\_Id}),
\item
   The type of raw data.
\end{itemize}

The template for a \emph{Composite} requires:

\begin{itemize}
\item
   The identifier of the new type (\emph{Dat\_Typ\_Id}),
\item
   A list of pairs (String, \emph{Dat\_Typ\_Id}).
\end{itemize}

A \emph{Composite} is created by providing a named list of \emph{MOON\_Data}
types, whose names must be unique within this list. These names form paths that
identify each \emph{MOON\_Data} within a \emph{Composite} \emph{MOON\_Data}.
The raw data is found in \emph{Singles}, which are the leaves of the trees that
make up the \emph{Composites}. In the end, we retrieve the structure of a file
system where \emph{Composites} are directories and \emph{Singles} are files.

This way of assembling data gives instances of \emph{MOON\_Data} the following
introspective properties:

\begin{itemize}
\item
   It is possible to know whether a \emph{MOON\_Data} is \emph{Single} or
   \emph{Composite},
\item
   It is possible to know whether a \emph{MOON\_Data} contains a given path
   and, if so, to access it,
\item
   In the case of a \emph{Single}, it is possible to know the type of its raw
   data and to access it,
\item
   Each path is associated with an access right, \emph{Read\_Only}, set to
   \emph{False} by default,
\item
   It is possible to set \emph{Read\_Only} to \emph{True}, with no possibility
   of returning it to \emph{False}.
\end{itemize}

Paths allow for generic introspection of all data. Access is controlled for
writing with the \emph{Read\_Only} property associated with each path. This
allows certain values to be fixed while leaving others free within the same
\emph{MOON\_Data} instance. The asymmetry of the \emph{Read\_Only} setting also
allows for step-by-step configuration through successive specifications and
parameterizations.


\subsection{Elementary Medical Specifications}
\label{elementary-medical-specifications}

\emph{This section presents how medical experts describe their elementary
activities by specifying types of Result and Prescription. More
complex processes expressed by Protocols are not covered here.}

Types \emph{Result} and \emph{Prescription} translate elementary medical
activities; medical pathways are sequences of instances of these
\emph{Record\_Elements}.

A new \emph{Record\_Element} type is created using a class template to which
the necessary information is provided.

The \emph{Result} template requires the following information:

\begin{itemize}
\item
   \emph{Elt\_Typ\_Id}: identifier of this new type of \emph{Record\_Element},
\item
   \emph{Category}: \emph{Exam}, \emph{Treatment}, or \emph{Inference}
   (\emph{Program} is forbidden),
\item
   \emph{Dat\_Typ\_Id}: identifier of the type of \emph{MOON\_Data} embedded in
   this \emph{Result}.
\end{itemize}

The \emph{Prescription} template requires the following information:

\begin{itemize}
\item
   \emph{Elt\_Typ\_Id}: identifier of this new type of \emph{Record\_Element},
\item
   \emph{Elt\_Typ\_Id}: identifier of the type of \emph{Result} produced by
   this \emph{Prescription},
\item
   \emph{Dat\_Typ\_Id}: identifier of the type of \emph{MOON\_Data} embedded in
   this \emph{Prescription} (type of the prescription's parameter),
\item
   \emph{Default\_Param}: default parameterization with a \emph{MOON\_Data}
   instance of type \emph{Dat\_Typ\_Id},
\item
   \emph{Dependencies}: list of \emph{Dependency} for this \emph{Prescription}.
\end{itemize}

The rules associated with the templates are as follows:

\begin{itemize}
\item
   All types must be identified,
\item
   Referencing a type by its identifier assumes that this type exists (i.e., it
   has already been created),
\item
   The \emph{Program} category is forbidden (all \emph{Record\_Elements} of
   this category are already defined in the base model),
\item
   A \emph{Prescription} is of the same category as its \emph{Result},
\item
   Multiple \emph{Prescriptions} can produce the same type of \emph{Result},
\item
   An \emph{Exam Prescription} does not admit any \emph{Dependency},
\item
   An \emph{Inference Prescription} has at least one \emph{Dependency},
\item
   A \emph{Treatment Prescription} can have \emph{Dependencies}.
\end{itemize}

The two fundamental concepts of the model, \emph{Prescription} and
\emph{Result}, embed \emph{MOON\_Data} as parameters or results.

To develop a set of \emph{Record\_Elements} covering a domain, medical experts
must consider their elementary activities in terms of \emph{Prescriptions} and
\emph{Results}, and/or reconsider their raw data under the criteria of
\emph{MOON\_Data} and these two templates:

\begin{itemize}
\item
   Differentiate everything that relates to prescription from what relates to
   the result,
\item
   For a \emph{Result}:

   \begin{itemize}
   \item
      Build the appropriate \emph{MOON\_Data} from the raw data,
   \item
      Associate the correct category: is it an observation, a treatment, or
      purely an idea?
   \end{itemize}
\item
   For a \emph{Prescription}:

   \begin{itemize}
   \item
      Identify the \emph{Result},
   \item
      Ask how this \emph{Result} is built. What information is needed, and what
      is the nature of each: parameter or preliminary result?

      \begin{itemize}
      \item
         For parameters:

         \begin{itemize}
         \item
            Group them in the form of a \emph{MOON\_Data},
         \item
            What are the default values for this parameterization?
         \item
            And which values are fixed, which are free?
         \end{itemize}
      \item
         For dependencies:

         \begin{itemize}
         \item
            Which types does the prescription depend on?
         \item
            How many preliminary results are needed?
         \item
            And what age should they not exceed?
         \end{itemize}
      \end{itemize}
   \end{itemize}
\end{itemize}

A registry groups all created \emph{Record\_Element} types. This registry
allows for the validation or invalidation of the creation of new types. The
registry can be queried; it provides access to a wide variety of information, for
example, knowing which exams provide information on a given pathology. It forms
the knowledge base that allows for the construction of shared and universally
understood pathways.

A set of \emph{Record\_Elements} that more or less fully covers the elementary
activities of a specialized medical domain constitutes a medical model of that
domain. Two medical models of different specialties can coexist effectively if
the elements they have in common are redefined in a more general model. The
union of the three models then covers a broader domain.


\subsection{Pathway Actors}
\label{pathway-actors}

\emph{This section introduces the concept of Service, making it possible
to identify the participants and define their roles.}


\subsubsection{Service, Structure, and Identifier}
\label{service-structure-and-identifier}

A pathway can involve a wide variety of actors. They are all identified by
their name and role as a \emph{Service}. \emph{Service} is an abstraction that
can take three forms: \emph{Place}, \emph{System} and \emph{Person}. For
example, medical practices, hospitals, and hospital departments are
\emph{Places}; equipment and software are \emph{Systems}; and doctors, nurses
are \emph{Persons}.

A \emph{Place} can have sub-services, unlike a \emph{Person} or a
\emph{System}. This makes it possible to structurally represent an
organization. A \emph{Service} identifier is a path that locates it within the
organization; it can be expressed absolutely or relative to its \emph{Place}.

An example of service identifiers for a fictional imaging center, which has 2
imagers, 2 operators, and 2 softwares is shown on Table \ref{table:knock}.

\begin{table}[htbp]
\caption{\label{table:knock} Exemple of \emph{Service} identifiers.}
\begin{center}
\begin{tabular}{ll}
\hline
\textbf{Path - Identifier} & \textbf{Service Type} \\
\hline
Knock\_Imaging & Place \\
Knock\_Imaging/Dr\_Knock & Person \\
Knock\_Imaging/Img1 & Place \\
Knock\_Imaging/Img1/Imaging\_Device & System \\
Knock\_Imaging/Img1/Operator & Person \\
Knock\_Imaging/Img2 & Place \\
Knock\_Imaging/Img2/Imaging\_Device & System \\
Knock\_Imaging/Img2/Operator & Person \\
Knock\_Imaging/Info & Place \\
Knock\_Imaging/Info/Software1 & System \\
Knock\_Imaging/Info/Software2 & System \\
\hline
\end{tabular}
\end{center}
\end{table}

The same example could have been handled differently; the important thing is
that the naming scheme follows the logic of the service.

A \emph{Service} does not represent the actor but the set of roles it is
expected to fulfill within its service, in other words, everything that is
expected from that service. Thus, \emph{Knock\_Imaging/Dr\_Knock} does not identify the
person but their activity within this structure. If Dr.~Knock also works in
another facility, this is a different \emph{Person} with a different identifier
and different roles.

\ifthenelse{\equal{\MoonArticleMode}{IEEEJBHI}}
{\textit{Case of patients}}
{\paragraph{Case of patients}}
\label{patient-service}
the patient also plays roles in its pathway but should be handled specifically
considering their special place. In MOON, the \emph{Patient\_Profile} lists
actions the patient could have as a \emph{Person} and also as a \emph{System}
taking into account any apparatus they could carry with them. This special
\emph{Place} contains one \emph{Person} and only one, and several \emph{Places}
and \emph{Systems} to organize and represent the equipments they may carry such
as pacemaker, glocumeter, or smartwatch.


\subsubsection{Service Offer, Prescriber, and Operator}
\label{service-offer-prescriber-and-operator}

All pathway actors are identified as \emph{Service}. A \emph{Service} is
characterized by its offer, namely the list of all roles it can fulfill in a
medical pathway. The offer consists of instances of \emph{Protocols} and
\emph{Prescriptions} available for scheduling. Each instance is associated with
a certain amount of operational information that formalizes its use, such as
availability (see the following section \ref{service-contract}). The
\emph{Service} thus publishes the roles it can fulfill in the form of service
contracts. Each instance of the offer references the \emph{Service} by its
identifier. Thus, the \emph{Prescriptions} and \emph{Protocols} of a pathway
clearly identify the actor who handles them.

The actor who schedules an instance proposed by a service is referenced in
the \emph{Program\_Prescription} used for this purpose.

Finally, the pathway identifies the actors of all pathway steps, specifying
whether they are \emph{prescribers} or \emph{operators}, i.e., who requests and
who performs. \emph{Prescribers} are identified by \emph{Prescriptions} of the
\emph{Program} category; \emph{operators} are identified by
\emph{Prescriptions} of the other categories.

Here are two examples:

\textbf{Individual Prescriptions} - A general practitioner schedules various
\emph{Prescriptions} for exams proposed individually by different
\emph{Services}, then schedules the next follow-up visit in their office as soon
as possible:

\begin{itemize}
\item
   The \emph{prescriber} of the exams and the next visit is the general
   practitioner,
\item
   The \emph{operator} of each exam is the relevant service,
\item
   The \emph{operator} of the next visit is the general practitioner.
\end{itemize}

\textbf{Prescriptions in a Protocol} - This time, all exams are in a protocol
from the same service; the general practitioner schedules this \emph{Protocol}
and then the visit:

\begin{itemize}
\item
   The \emph{prescriber} of the protocol and the next visit is the general
   practitioner,
\item
   The \emph{prescriber} of the exams is the service,
\item
   The \emph{operator} of the exams is the service,
\item
   The \emph{operator} of the next visit is the general practitioner.
\end{itemize}


\subsubsection{Service Contract}
\label{service-contract}

Each instance of \emph{Prescription} or \emph{Protocol} in the service offer
must be carefully parameterized to constitute a true service contract in which
all roles appear unambiguously and are ready for use.

Like the specification of types, the service offer excludes
\emph{Prescriptions} of the \emph{Program} category.
These prescriptions are managed internally by the \emph{Record} concept, which
encapsulates the pathway; this concept is presented later in the article.

\textbf{Parameterization of a \emph{Prescription}} - Each type of
\emph{Prescription} has a parameter in the form of \emph{MOON\_Data}
(\ref{raw-data-packaging}), whose value was chosen by the expert who specified
the \emph{Prescription} type. The expert has left open certain paths of that
parameter for writing access. The freedom left by the type specifier is
reinvested in the same way by the \emph{Service}, which chooses new values in
this space and closes it as it wishes. Thus, it introduces into its offer an
instance of \emph{Prescription} that will be prescribed at the desired value,
with or without room to maneuver for the prescriber.

To offer a type of \emph{Prescription} with multiple parameterization
possibilities, the service can choose to leave certain paths open for the
prescriber and/or offer several closed instances with different settings.

\textbf{Parameterization of a \emph{Protocol}} - The parameterization of a
\emph{Protocol} follows the same logic as that of a \emph{Prescription}, based
on \emph{MOON\_Data} access. However, it is important to note that a
\emph{Service} can subcontract \emph{Prescriptions} and \emph{Protocols}
through its \emph{Protocols}. Thus, to parameterize a \emph{Protocol}, it must
be specified which \emph{Service} operates each of its components.

\ifthenelse{\equal{\MoonArticleMode}{IEEEJBHI}}
{}
{\paragraph{}}
\label{p:timing-presc-and-res}
\textbf{Timing of a \emph{Prescription} and \emph{Result} Dates} - A type of
\emph{Prescription} does not specify the time required to operate or to deliver
the result. Two different \emph{Services} do not necessarily operate the same
\emph{Prescription} with the same delays, and the same \emph{Service} may take
more or less time depending on the parameterization of a \emph{Prescription}.
Therefore, it is up to the \emph{Service} to specify this information and make
it visible in the contract. For each instance of \emph{Prescription} in the
offer, the \emph{Service} must indicate three durations:

\begin{itemize}
\item
   \textbf{\emph{Duration}}: estimated total duration of the action,
\item
   \textbf{\emph{Production\_Delay}}: duration of result production, less
   than or equal to \emph{Duration},
\item
   \textbf{\emph{Delivery\_Delay}}: maximum duration of result delivery
   following the action.
\end{itemize}

These three durations translate into the pathway as dates associated with each
\textbf{\emph{Result}}:

\begin{itemize}
\item
   \textbf{\emph{Begin\_Date}}: start date of the action,
\item
   \textbf{\emph{Production\_Date}}: date of result production,
\item
   \textbf{\emph{End\_Date}}: end date of the action,
\item
   \textbf{\emph{Delivery\_Date}}: date of delivery of the \emph{Result} value,
\item
   \textbf{\emph{Value\_Date}}: date of the \emph{Result} value.
\end{itemize}

With:
\ifthenelse{\equal{\MoonArticleMode}{IEEEJBHI}}
{
\begin{IEEEeqnarray*}{rcccl}
End\_Date &=& Begin\_Date &+& Duration \\
Production\_Date &=& Begin\_Date &+& Production\_Delay \\
Delivery\_Date &=& End\_Date &+& Delivery\_Delay
\end{IEEEeqnarray*}
}
{
$$
\begin{array}{rclcl}
End\_Date &=& Begin\_Date &+& Duration \\
Production\_Date &=& Begin\_Date &+& Production\_Delay \\
Delivery\_Date &=& End\_Date &+& Delivery\_Delay
\end{array}
$$
}

For an \emph{Exam} or \emph{Treatment}, the \emph{Value\_Date} is equal to the
\emph{Production\_Date}. For an
\emph{Inference}, the \emph{Value\_Date} of the \emph{Result} is the last
\emph{Value\_Date} of its \emph{Pre\_Results}.

For an \emph{Exam} or \emph{Treatment}, the patient is occupied between
\emph{Begin\_Date} and \emph{End\_Date}; it is not possible to prescribe
another \emph{Exam} or \emph{Treatment} during this time.

\textbf{Place Exclusivity} - A \emph{Service} can obtain the exclusivity of
the patient for a given duration. This prevents external scheduling from
interfering during this time period.

This possibility is limited to services of type \textbf{\emph{Place}}. Two
additional pieces of information are added to any instance of
\emph{Prescription} or \emph{Protocol} for this type of \emph{Service}:

\begin{itemize}
\item
   \textbf{\emph{On\_Site}}: boolean activating place exclusivity,
\item
   \textbf{\emph{Stay}}: duration of stay in the \emph{Service}.
\end{itemize}

An offer instance of a \emph{Place} with this option activated imposes the
patient's presence in this \emph{Place} throughout this time. During this
period, it is no longer possible to schedule an \emph{Exam} or a
\emph{Treatment} operated outside this service or one of its sub-services
(unless from the patient themselves as a \emph{Patient\_Profile}). Conversely,
this instance cannot be prescribed if an \emph{Exam} or \emph{Treatment}
outside the \emph{Place} is already scheduled during the considered time
period.

The main use of this option is to enable \emph{Appointments}. A \emph{Place}
that offers an empty \emph{Protocol} associated with this option offers nothing
else than an \emph{Appointment}. Whether it is a general practitioner or a
hospital, the patient will be entirely available so that the service in
question and its sub-services can perform all necessary scheduling during the
considered period.

\textbf{Availability} - The final aspect of the contract concerns the service's
availability. Each instance of offer is accompanied by its \emph{opening
hours}. The \emph{Service} indicates whether a given date is possible for an
offer instance or what the next available date is.


\subsubsection{Referencing Service Offers}
\label{referencing-service-offers}

Like the types of \emph{Record\_Elements} and \emph{Protocols}, the instances
of offers must be published to be available. The base model does not specify
the aspects of referencing but simply imposes:

\begin{itemize}
\item
   That a specifying expert has the means to publish a new type,
\item
   That a service also as the means to publish its offer,
\item
   And that a prescriber has access to these publications to inform themselve,
   document, compare, and ultimately prescribe.
\end{itemize}

The model does not impose a solution but allows for the definition of all these
means.

The base model allows for imagining a variety of tools related to referencing,
enabling queries on the registries of types and offers, highlighting
differences between two types or two offers, or searching for an offer by
geographical or temporal proximity.

The hierarchical structure of \emph{Service} allows for adapting the
visibility of offers. Internal and restricted publication for the use of a
\emph{Service} allows for distributing roles within the organization without
directly offering them to the public. Thus, in the previous example of
Dr.~Knock (Table \ref{table:knock}), the \emph{Service} \emph{Knock\_Imaging} propose
a set of public offers (\emph{Prescriptions} and/or \emph{Protocols}) that it
can operate. Internally these public offers refer to offer instances from its
sub-services from a private registry. The \emph{Service} \emph{Imagerie\_Knock}
is the public offer interface and encapsulates the offers of the sub-services,
which are invisible to the public.

\ifthenelse{\equal{\MoonArticleMode}{IEEEJBHI}}
{\textit{Case of patients}}
{\paragraph{Case of patients}}
\label{patient-profile}
A \emph{Patient\_Profile} does not identify a given patient, it is a public
profile used to specify what a patient can do and what is expected from
them. In other words, it is not an offer from the patient but specific
responsibilities given to them. A profile is not permanently attached to the
patient but can be assigned temporarily. More over a patient can be assigned
multiple profiles to take into account different roles at the same time.

These profiles can be published by a medical community sharing the same
practices, or by the manufacurer of a specific device. For example, in the case
of diabetes, one can imagine several usual profiles of patients, and specific
profiles associated to different types of devices such as automatic glucose
meters.


\subsection{Record}
\label{record}

\emph{This section presents the concept of record, which controls access
patient's medical pathway and current profiles.}

The \textbf{\emph{Record}} is the concept used to represent and manage the
medical pathway and the list of \emph{Patient\_Profiles}. It encapsulates these
two critical items so that all their modifications can be strictly controlled.
They are done through the \emph{Record} interface, whose functionalities are
presented in five aspects:

\begin{itemize}
\item
   Consult and explore the pathway,
\item
   Schedule actions,
\item
   Advance through time,
\item
   Fill in the value of a result,
\item
   Consult or modify the list of \emph{Patient\_Profiles} assigned to the patient.
\end{itemize}

These five aspects are presented point by point in the following sections.


\subsubsection{Exploring the Pathway}
\label{exploring-the-pathway}

The record contains all instances of \emph{Record\_Element} plus a timeline
that aggregates all the dates of the pathway. The instances of
\emph{Record\_Elements} are connected to each other by their scheduling and
dependency relationships. The timeline groups the five dates of each
\emph{Result} as well as the exclusivity periods of \emph{Place}. The
present is identified by its date and its position relative to the timeline
dates. The whole can be queried by combining several aspects of the pathway:

\begin{itemize}
\item
   Types or values of dates,
\item
   Availability,
\item
   Decision chains,
\item
   Dependency chains,
\item
   Types or values of \emph{Results},
\item
   Types or values of \emph{Prescriptions},
\item
   Types or values of \emph{Protocols},
\item
   Paths, types, or values of \emph{MOON\_Data} associated with components,
\item
   Identifiers of prescribers or operators,
\item
   etc.
\end{itemize}

Based on these queries, it is straightforward to develop a wide variety of tools to
answer a variety of medical and operational questions:

\begin{itemize}
\item
   What is the next date in the timeline, and what does it consist of?
\item
   Is there an exam of type X in the pathway?
\item
   What is the date of the last treatment of type X completed?
\item
   What is the value of the last exam of type X completed?
\item
   What are the dates and values of all past exams of type X?
\item
   Is a treatment of type X planned?
\item
   What are the parameters of this treatment?
\item
   On which results is a particular inference result based?
\item
   By whom or what was this inference operated?
\item
   Is this result used as a premise, and by which results?
\item
   What is the decision chain of this result?
\item
   What are the preliminary results compatible with a given dependency at a
   given date?
\item
   What are the \emph{Protocols} in this decision chain?
\item
   Who are the actors in this decision chain?
\item
   What is the production chain of this result?
\item
   Who are the actors in this production chain?
\item
   Is the patient available between these two dates?
\item
   When was a given exam be scheduled?
\item
   When will a given exam be available?
\item
   When was a given exam performed?
\item
   How much time has passed between these two exams?
\item
   What is the impact of a given unscheduling?
\item
  etc.
\end{itemize}


\subsubsection{Scheduling Actions}
\label{scheduling-actions}

Unlike queries, scheduling or unscheduling anything modifies the pathway. These
modifications are prescribed using \emph{Prescriptions} whose category
\emph{Program} is forbidden to specifiers, services, and prescribers to ensure
the consistency of the pathway. Its implementation is encapsulated by the
\emph{Record} through three properties:

\begin{itemize}
\item
   \textbf{\emph{Prescribe}}: to schedule an instance of \emph{Protocol} from a
   future date,
\item
   \textbf{\emph{Prescribe\_Single\_Asap}}: to schedule an instance of
   \emph{Prescription} as soon as possible,
\item
   \textbf{\emph{Abort\_Program}}: to interrupt an ongoing program.
\end{itemize}

All three properties act in the present. They add a pair of
\emph{Prescription/Result} of type \emph{Program}; the dates of the
\emph{Program\_Result} are at the current date.

\vskip 3mm
\textbf{\emph{Prescribe}} takes the following parameters:

\begin{itemize}
\item
   The identifier of the prescriber,
\item
   A future start date for scheduling,
\item
   The patient's availability,
\item
   The identifier of the service that will operate the \emph{Protocol},
\item
   The identifier of the \emph{Protocol} offer within the service,
\item
   A list of \emph{Pre\_Results} satisfying the dependencies of this
   \emph{Protocol},
\item
   Optionally, a list of \emph{Path/Moon\_Data} pairs satisfying free access to
   the parameterization of this \emph{Protocol}.
\end{itemize}

\emph{Prescribe} schedules the service's protocol, taking into account the
patient's and services' availability. The prescriber can parameterize the
protocol within the limits set by the service and must indicate the preliminary
results of the pathway that the protocol requires.

Technically, this property adds to the pathway a \emph{Prescription} of type
\emph{Program\_Prescription}, whose parameter is a copy of the offer instance
altered by the \emph{Pre\_Results} and any reparameterization. It also adds a
\emph{Result} of type \emph{Program\_Data}, which includes the
\emph{Prescriptions} of the \emph{Protocol} along with their scheduling dates.
Each of these \emph{Prescriptions} is added to the \emph{Record}, as well as a
\emph{Result} per date (\emph{Begin\_Date}). In case of scheduling
impossibility, the \emph{Program\_Data} contains an empty program associated
with a message indicating the reason for the failure.

Note: If the \emph{Protocol} contains other \emph{Protocols}, then each of them
becomes the parameter of a \emph{Program\_Prescription} of the
\emph{Program\_Data} (recursive process). If the scheduling of an internal
\emph{Protocol} fails, it causes the failure of its enclosing protocol; the
error is reported step by step, limiting the addition of a single
\emph{Program\_Data} to the \emph{Record}.

\vskip 3mm
\textbf{\emph{Prescribe\_Single\_Asap}} takes the following parameters:

\begin{itemize}
\item
   The identifier of the prescriber,
\item
   The patient's availability (for an \emph{Exam} or \emph{Treatment}),
\item
   The identifier of the service that will operate the \emph{Prescription},
\item
   The identifier of the \emph{Prescription} offer within the service,
\item
   A list of \emph{Pre-Results} satisfying the dependencies of this
   \emph{Prescription},
\item
   Optionally, a list of \emph{Path/Moon\_Data} pairs satisfying free access to
   the parameterization of this \emph{Prescription}.
\end{itemize}

\emph{Prescribe\_Single\_Asap} is the equivalent of \emph{Prescribe} for a
single \emph{Prescription} and is expressed in much the same way. The earliest
date is calculated based on the patient's and service's availability. This
property encapsulates the call to \emph{Prescribe} for a
\emph{Protocol\_Program} consisting simply of the \emph{Prescription} and the
calculated date.

\vskip 3mm
\textbf{\emph{Abort\_Program}} takes the following parameters:

\begin{itemize}
\item
   The identifier of the prescriber (the stopper),
\item
   The \emph{Program\_Result} to be stopped.
\end{itemize}

\emph{Abort\_Program} allows stopping an ongoing program. All future
\emph{Results} of this program are marked as \emph{Aborted}, and their values
will never be filled in. Note that a result whose preliminary results include a
result marked as \emph{Aborted} must also be marked as such. Thus, stopping a
program can trigger the stopping of other programs in a cascade.

Technically, \emph{Abort\_Program} adds an \emph{Abort\_Prescription} of
category \emph{Program}, whose parameter is the \emph{Program\_Result} to be
stopped. The data of the corresponding \emph{Abort\_Result} provides
information on the cascade of aborted results. Each of these results is marked
with \emph{Aborted} status and with a reference to the \emph{Abort\_Result}.


\subsubsection{Advancing Through Time}
\label{advancing-through-time}

All dates of the \emph{Results} mark the timeline of the pathway; one can
consider the present as a cursor on this timeline. Two methods allow moving
along the time scale:

\begin{itemize}
\item
   \textbf{\emph{To\_Date}}: takes a date as a parameter and advances the
   present to that date; it must be in the future without exceeding the next
   milestone,
\item
   \textbf{\emph{To\_Next\_Event}}: advances the present to the next milestone.
\end{itemize}


\subsubsection{Filling in the Value of a Result}
\label{filling-in-the-value-of-a-result}

When the present is on a \emph{Production\_Date}, the present is locked on this
milestone until the value of the corresponding \emph{Result} is filled in. The
type of the prescription, the value of its parameters, and all preliminary
results are available to fill in the value of this \emph{Result}. Additionally,
only the operator identified by the operator's identifier can set the
\emph{Result} data value whose type has to be the data type of the
\emph{Result}'s data indicated in the \emph{Result} specification.

\subsubsection{Consult or modify the list of assigned \emph{Patient\_Profiles}}
\label{patient-profiles-list}
The roles that a patient can take in their journey are defined in the different
\emph{Patient\_Profiles} assigned to them.  The overall operations of these
\emph{Patient\_Profiles} represents all the responsabilities entrusted to that
patient, as a \emph{Person} or as a \emph{System}. It can be simple things like
taking a pill or complex things like controlling an insulin pump.
Before to prescribe any operation involving the patient, one must ensure that
the patient is able to perform the tasks that will be assigned to them, namely
that they have the consirered profile in their \emph{Record}, but also the
skills and the devices corresponding to the profile. For example, a
diabetologist who equips a patient with an automatic glucometer must add the
appropriate profile to its list after having taught them how to use it. In
the same way, when an antibiotic treatment ends, the corresponding profile can
be removed.

The \emph{Record} allows to edit this list, adding or remove profiles to the
patient.  Access at these modifications could be restricted to specific actors
(e.g., physicians) following a policy defined by the authority of the overall
system.

\newpage
\section{Uses of MOON}
\label{uses-of-moon}

\emph{This section presents the MOON model from the perspective of its use.}


\subsection{Three Phases of Use, Three Types of Users}
\label{three-phases-of-use-three-types-of-users}

The use of MOON is divided into three successive phases, each depending on the
previous one:

\begin{enumerate}
\def\labelenumi{\arabic{enumi}.}
\item
   Specification of the \textbf{medical model} (\ref{elementary-medical-specifications}),
\item
   Specification of \textbf{service offers} (\ref{service-contract}),
\item
   Use of \textbf{records} (\ref{record}).
\end{enumerate}

These three phases clearly separate users according to their responsibilities:

\begin{enumerate}
\def\labelenumi{\arabic{enumi}.}
\item
   Medical experts:

   \begin{itemize}
   \item
      Represent their profession and their domain of expertise,
   \item
      Specify their elementary activities from a data perspective,
   \item
      Specify scenarios based on elementary activities,
   \item
      Thus propose a medical model of their domain.
   \end{itemize}
\item
   Services:

   \begin{itemize}
   \item
      Are responsible for operating the elements of the medical model,
   \item
      Clearly and precisely publish which activities and scenarios they propose
      to operate.
   \end{itemize}
\item
   Record actors:

   \begin{itemize}
   \item
      Are the patient and all stakeholders,
   \item
      Take turns being prescribers or operators,
   \item
      Draw from service offers as prescribers,
   \item
      And fill in the value of a result as operators when it is their turn to do so.
   \end{itemize}
\end{enumerate}


\subsection{Formalism and Specifications}
\label{formalism-and-specifications}

MOON requires defining everything that can be defined as early as possible:
first, what the medical practice is, then who is responsible for which step
along medical pathway. This involves formalizing medical activities and service
offers. A few people with a high level of expertise dedicate time, rigor, and
precision so that everyone can subsequently benefit from this effort in using
extremely efficient records.

The formalism of activities and offers makes it possible to structure the
medical pathway systematically. The concrete scheduling of actions and their
reservations are ensured by the system, thus guaranteeing the completeness,
consistency, and validity of the pathway. The efficiency of such a process allows to gain a lot of time at every step.


\subsection{Uses of Specifications}
\label{uses-of-specifications}

Stakeholders are completely relieved of complex operations related to record
management. They focus on fundamental activities: observe, treat, deduce, and
decide. Furthermore, these activities are facilitated by the possibilities
offered respectively by \emph{the medical model, services, and the record}:

\begin{itemize}
\item
   \emph{The medical model} provides precise information on professional
   practices,
\item
   \emph{Services} provide information on everything that can be requested,
    where, and when,
\item
   \emph{The record}:

   \begin{itemize}
   \item
      Provides detailed information on the patient's complete history, past and
      future,
   \item
      Details all the responsabilities entrusted to the patient,
   \item
      Greatly facilitates scheduling with offers ready for use, and
      reservations made automatically at the same time,
   \item
      At the right time, the record controls each data entry, knowing the type
      of this data and who has to provide it.
   \end{itemize}
\end{itemize}

Let's take an example. A general practitioner schedules various exams,
appointments with specialists, and a follow-up visit in their office. All of this
is scheduled as soon as possible, within a given perimeter, taking into account
the patient's availability. After their appointment with the doctor, the patient
sees all the appointments scheduled, including the follow-up visit to that doctor. During each step, stakeholders have access to the pathway, perform the tasks assigned to them, and can also schedule
other possibilities. At the day of the follow-up visit, the general
practitioner has all the information they need, plus any additions to the
pathway, upcoming or already completed.

The pathway is shared by all actors, each benefiting from the work of others in
this unified collaborative framework. It is possible to schedule operations
involving multiple actors from different backgrounds without needing to gather
or inform them beforehand. General practitioners regain control over their
patients' pathways, and patients are completely relieved of any responsibility
regarding the management of their records.


\subsection{Operational Functionalities}
\label{operational-functionalities}

The pathway describes and sequences each step with precision. It can therefore
instruct and trigger operations. This opens up numerous operational
possibilities:

\begin{itemize}
\item
   Interact with the patient using a connected watch or smartphone:

   \begin{itemize}
   \item
      Supervision of medication intake (reminders and dating),
   \item
      Assistance with navigation within a hospital,
   \item
      Feedback on the progress of follow-ups,
   \item
      Dashboard, agenda, etc.
   \end{itemize}
\item
   Manage a service's workflow:

   \begin{itemize}
   \item
      Sequencing and triggering of automatic or interactive tasks,
   \item
      Automatic (batch tasks):
       \begin{itemize}
      \item
         Equipment control,
      \item
         Software control,
      \end{itemize}
   \item
      Interactive:

      \begin{itemize}
      \item
         Dashboard,
      \item
         Alarms,
      \item
        Data entry,
      \end{itemize}
   \item
      \emph{Unit\_Side} management:

      \begin{itemize}
      \item
         Administrative tasks, reception, payments, etc.,
      \item
         Logistical tasks, resource reservations, medications, meals, etc.
      \end{itemize}
   \end{itemize}
\item
   Monitor and alert:

   \begin{itemize}
   \item
      Individually for a given pathology,
   \item
      Collectively in case of an epidemic or emergency situation,
   \end{itemize}
\item
   Simulate the patient's medical pathway:

   \begin{itemize}
   \item
      Fill in a pathway using digital models of pathologies and treatments,
   \item
      Compare simulated pathways with real records,
   \item
      Propose new inferences based on predictive models,
   \item
      Medical education: simulate the doctor's situation facing the evolution
      of a virtual patient,
   \item
      Service simulation in routine situations: optimize configurations against a standard flow of virtual patients,
   \item
      Service simulation in extreme situations: test a configuration against specific flows of virtual patients (epidemic, natural disaster, war).
   \end{itemize}
\end{itemize}

Most of these new possibilities are on the service side and do not necessarily
relate to purely medical aspects. This is why the model extension provides a
\emph{Unit\_Side} category, in addition to \emph{Exam}, \emph{Treatment},
\emph{Inference}, and \emph{Program}, allowing for the addition of any
non-medical information useful for the service's operation to the pathway.
These additions, however, remain in the service's EMR and do not appear in the
shared EHR. This extension allows the service to also benefit from the MOON
model internally to manage its workflow according to its own considerations, in
continuity with the public EHR pathway.


\subsection{Secondary Uses of Records}
\label{secondary-uses-of-records-1}

There are no longer any obstacles to statistical studies. They can be
prospective, retrospective, or even both. Records are consistent, and the
information they contain is rich, unambiguous, valid, and accessible. On a
large scale, these statistical studies allow for:

\begin{itemize}
\item
   Highlighting correlations associated with rare diseases,
\item
   Highlighting the best responses to a given problem,
\item
   Highlighting the worst responses to a given problem,
\item
   Highlighting the lack of consideration for a given problem,
\item
   Highlighting the effectiveness or ineffectiveness of a treatment,
\item
   Highlighting the most effective and/or fastest and/or cheapest protocols,
\item
   Locating geographycally and/or in time the findings,
\item
   Early detection of a virus and monitoring its evolution,
\item
   Monitoring the introduction of a new medication and detecting any
   contraindications as early as possible,
\item
   Alerting about the risk of reaching the limit of a resource,
\item
   etc.
\end{itemize}

Secondary uses become very simple, with no issues of access or interpretation
of data. Mass exploration of records allows for advancing knowledge in medicine
and its management. Mass monitoring of records enables real-time supervision of
a large number of indicators, making health policies relevant, effective, and
agile.


\subsection{EHR or EMR}
\label{ehr-or-emr}

The use of EHRs is marginal compared to the multitude of EMRs. Because it is
difficult to impose a standard without providing direct benefits from its
use, MOON, with its unique functionalities, could help promote the use of EHRs.

It is possible to benefit from the operational advantages of MOON in a
perimeter where one has sufficient political and technical authority to ensure
its implementation. Here are a few examples:

\begin{itemize}
\item
   A hospital department with complex, frequently evolving processes could use
   MOON to express them in the form of protocols, simulate them, optimize
   them, manage them, and evolve them,
\item
   A service that wants to maximize interactions with the patient without
   generating a workload for staff can inform the patient about the progress
   of its pathway within the service, allow the patient to perform tasks
   interactively or by equiping it with automatic devices,
\item
   A country's health insurance system could start using MOON with the release
   of a new revolutionary but very expensive treatment that all patients
   demand. The paying entity could then require doctors to use MOON to
   prescribe this medication. It would receive direct feedback on a national
   scale and quickly know when to use this medication and when to do
   without. This entity would thus have the means to control public spending
   costs and evolve the prescription policy for this medication.
\end{itemize}

These three examples show relatively complex situations that are generic use
cases of the pathway and do not require specific development of the record.


\subsection{Formal Medical Language}
\label{formal-medical-language}

Medicine has an extremely rich vocabulary, allowing for the description of any
medical aspect with rigor and precision. However, it lacks an equivalent in
grammar and conjugation to articulate this vocabulary and describe a medical
pathway with the same level of rigor.

MOON is a kind of formal language that fills this gap through the abstraction
of medical logic. MOON expresses medicine through practices, roles, and
pathways. The abstraction describes the progression of activities in an
unambiguous and computationally interpretable manner.

MOON, as a language, can promote medical memory and communication in general.
The rigorous description of protocols and pathways must also find applications
in education and research.

Finally, it is also a response to a current trend that considers Artificial
Intelligence as a solution to all problems. Indeed, regarding medical records,
there is an growing tendency to hope that an AI will reconstruct the patient's
global pathway from information recovered in various EMRs. In other words, we
ask AI to guess what a doctor has deliberately thought. Why not ask the doctor
directly?

This is exactly what MOON does, while also allowing the doctor to express their
practices.

\newpage
\section{Graphical Illustrations}
\label{graphical-illustrations}

\begin{figure*}
\ifthenelse{\equal{\MoonArticleMode}{IEEEJBHI}}
 {
 \includegraphics[width=0.9\textwidth]{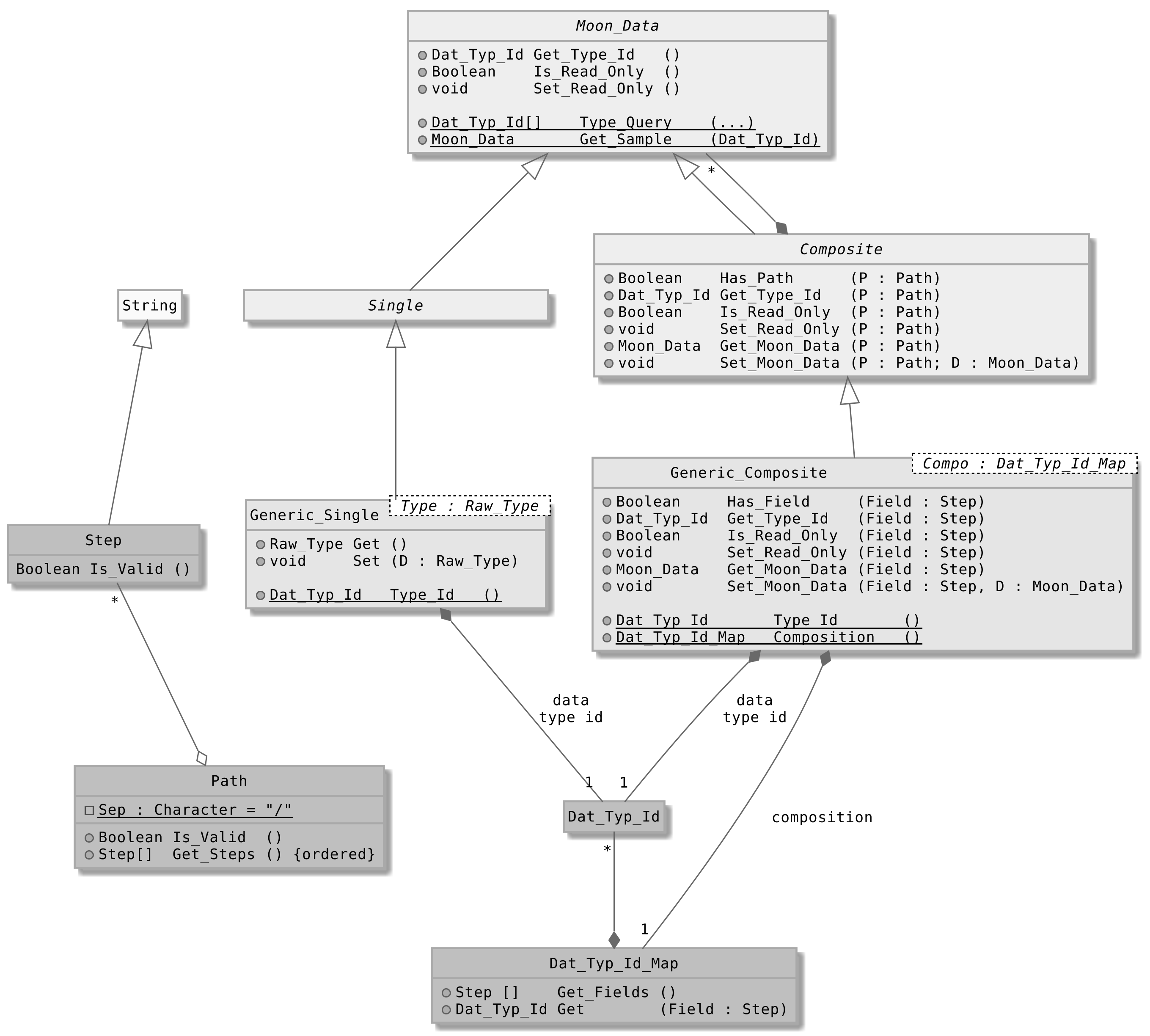}
 }
 {
 \includegraphics[width=0.9\textwidth]{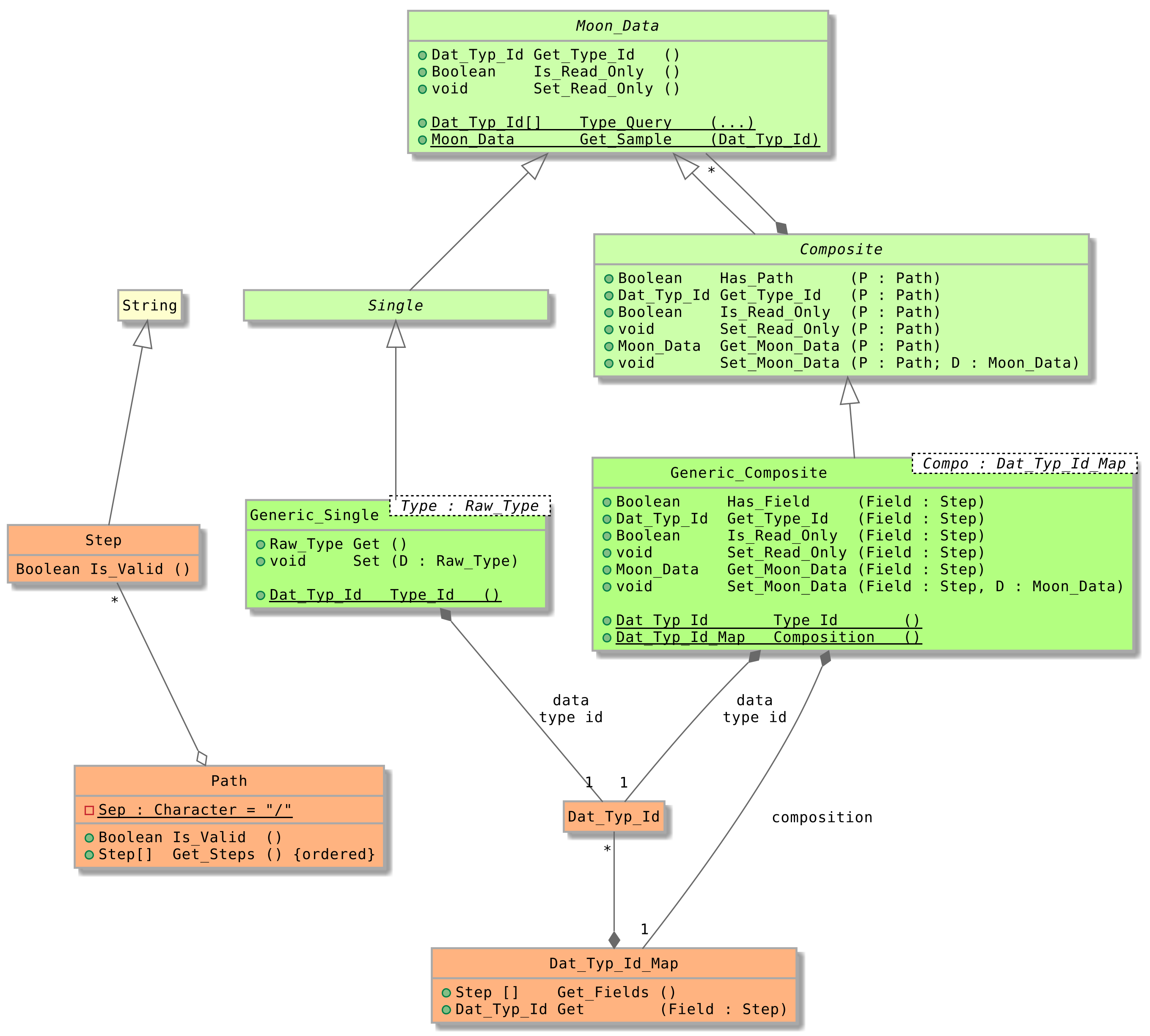}
 }
   \caption{UML class diagram of \emph{Moon\_Data} whose purpose is to
   systematize access to raw data. As explained in section
   \ref{raw-data-packaging}, raw data are embedded in the introspective
   abstract class \emph{MOON\_Data}. Each type of \emph{MOON\_Data} is
   identified by \texttt{Dat\_Typ\_Id}, its introspection is carried out from
   \texttt{Path}. With its two static methods, \emph{MOON\_Data} act as a
   factory based on \texttt{Dat\_Type\_Id} to query and get sample instances.
   \emph{MOON\_Data} is derived in two classes, \emph{Single} and \emph{Composite}.
   \emph{Single} encapsulates raw data while \emph{Composite} aggregates
   \emph{MOON\_Data}. The creation of a new type of \emph{MOON\_Data} is
   achieved by instantiating one of the two generic classes
   \emph{Generic\_Single} and \emph{Generic\_Composite}. The first create a new
   derived class of \emph{Single} by passing a raw data type to its formal
   parameter, the other create a new derived class of \emph{Composite} by
   passing a dictionnary of \texttt{Dat\_Typ\_Id}.}
 \label{f:MOONDataUML}
\end{figure*}

\begin{figure*}
\ifthenelse{\equal{\MoonArticleMode}{IEEEJBHI}}
 {
 \includegraphics[width=0.9\textwidth]{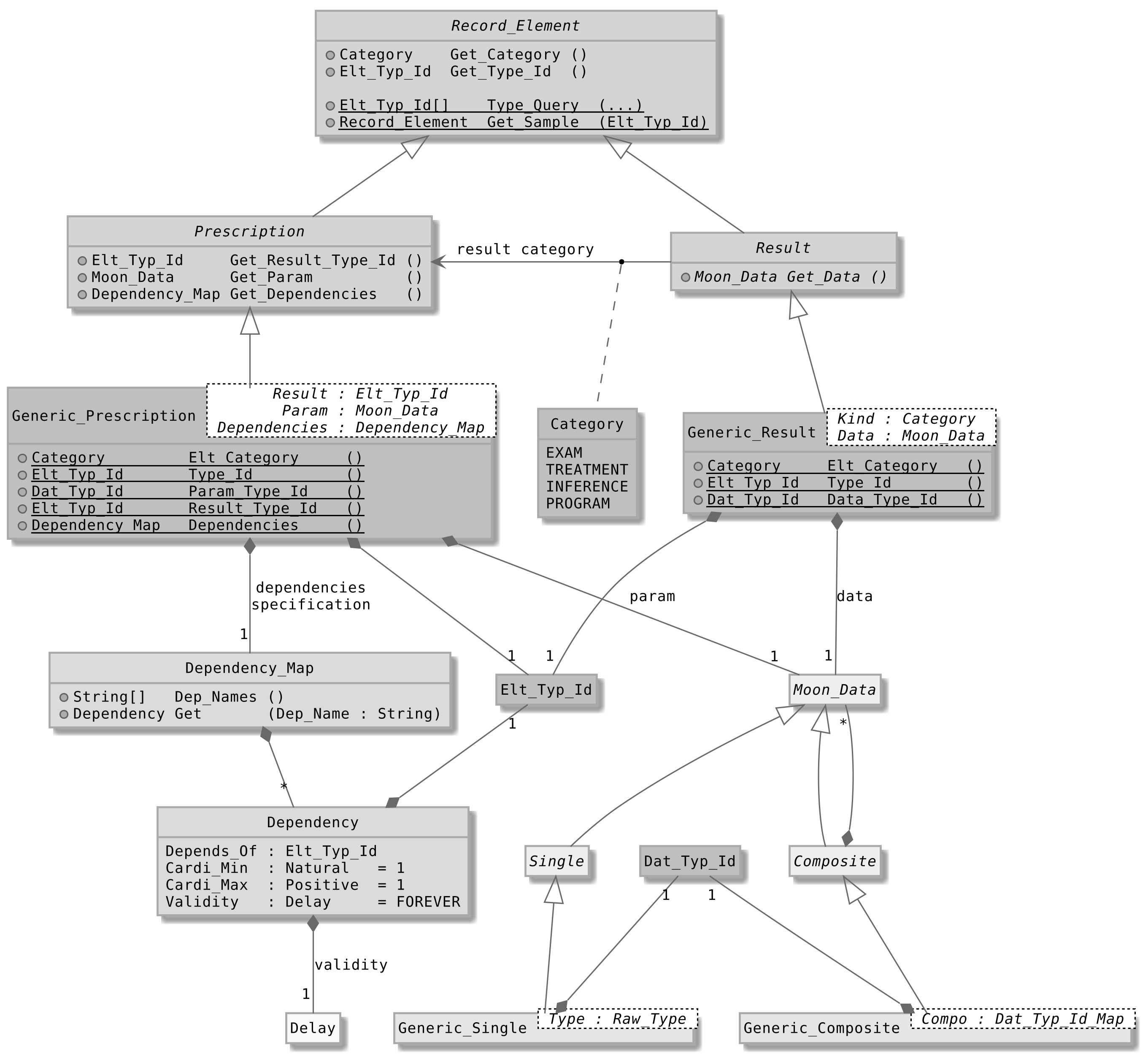}
 }
 {
 \includegraphics[width=0.9\textwidth]{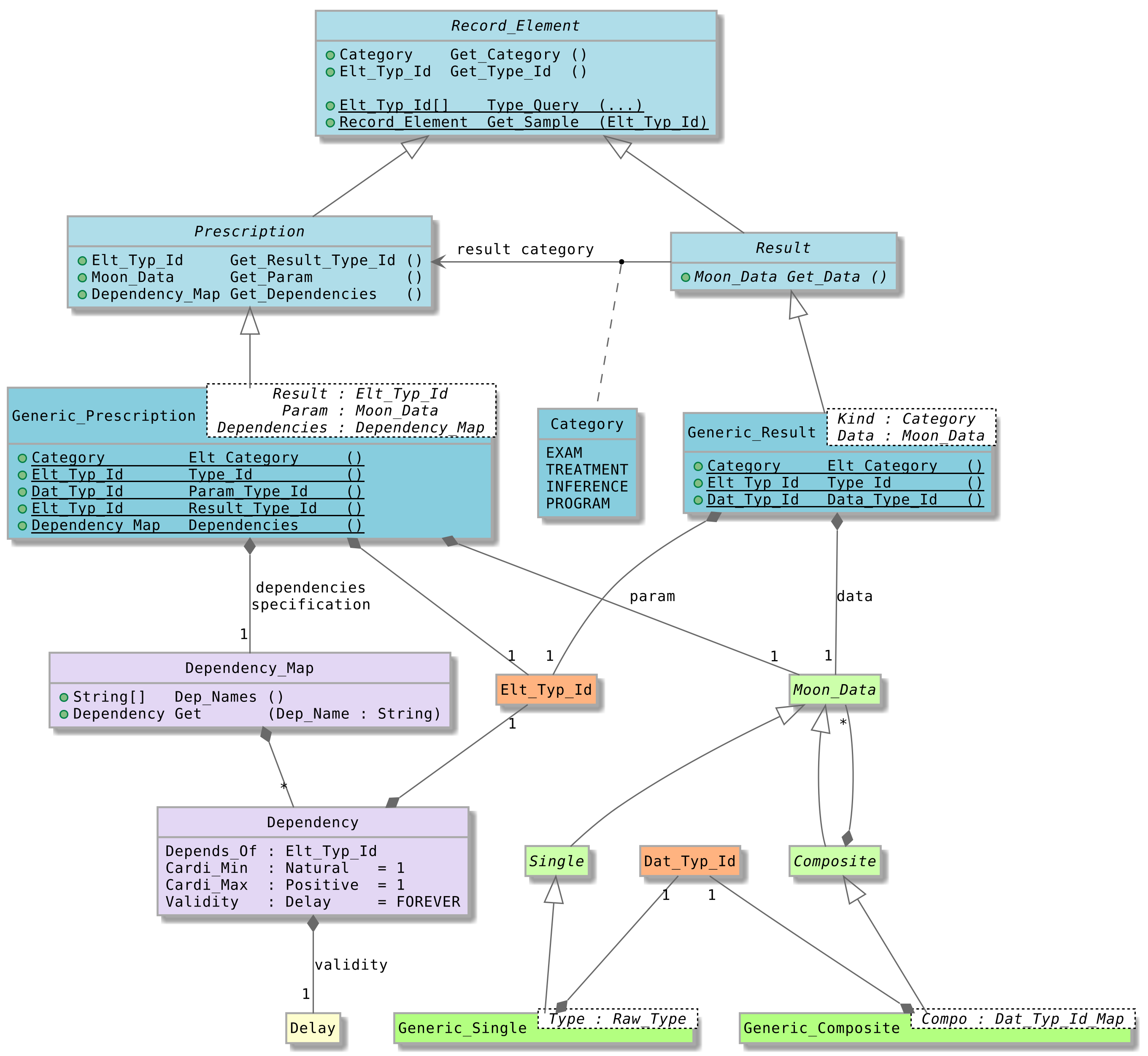}
 }
  \caption{UML class diagram of \emph{Record\_Element} introduced section
   \ref{scheduling}, which instances are the base components of a patient
   pathway. \emph{Record\_Element} is derived in two classes
   \emph{Prescription} and \emph{Result} which are (with \emph{Protocol}) the
   base components of a medical model created by experts of the considered
   domain. This diagram shows how to create these classes as explained in
   section \ref{elementary-medical-specifications}. As for \emph{MOON\_Data},
   new class of \emph{Record\_Element} are created by generic classes. A new
   class of \emph{Result} is created by passing a \emph{Category}
   (\ref{elementary-medical-specifications}) and a type of \emph{MOON\_Data},
   while a new class of \emph{Prescription} is created by passing three formal
   parameters to specify the type of its \emph{Result}, the type of its
   parameter and their dependencies. The new \emph{Prescription} take the
   \emph{Category} of its \emph{Result} and their dependencies are defined by a
   dictionnary of \emph{Dependency} (\ref{dependencies}).}
 \label{f:RecElementsUML}
\end{figure*}

\begin{figure*}
\ifthenelse{\equal{\MoonArticleMode}{IEEEJBHI}}
 {
 \includegraphics[width=0.9\textwidth]{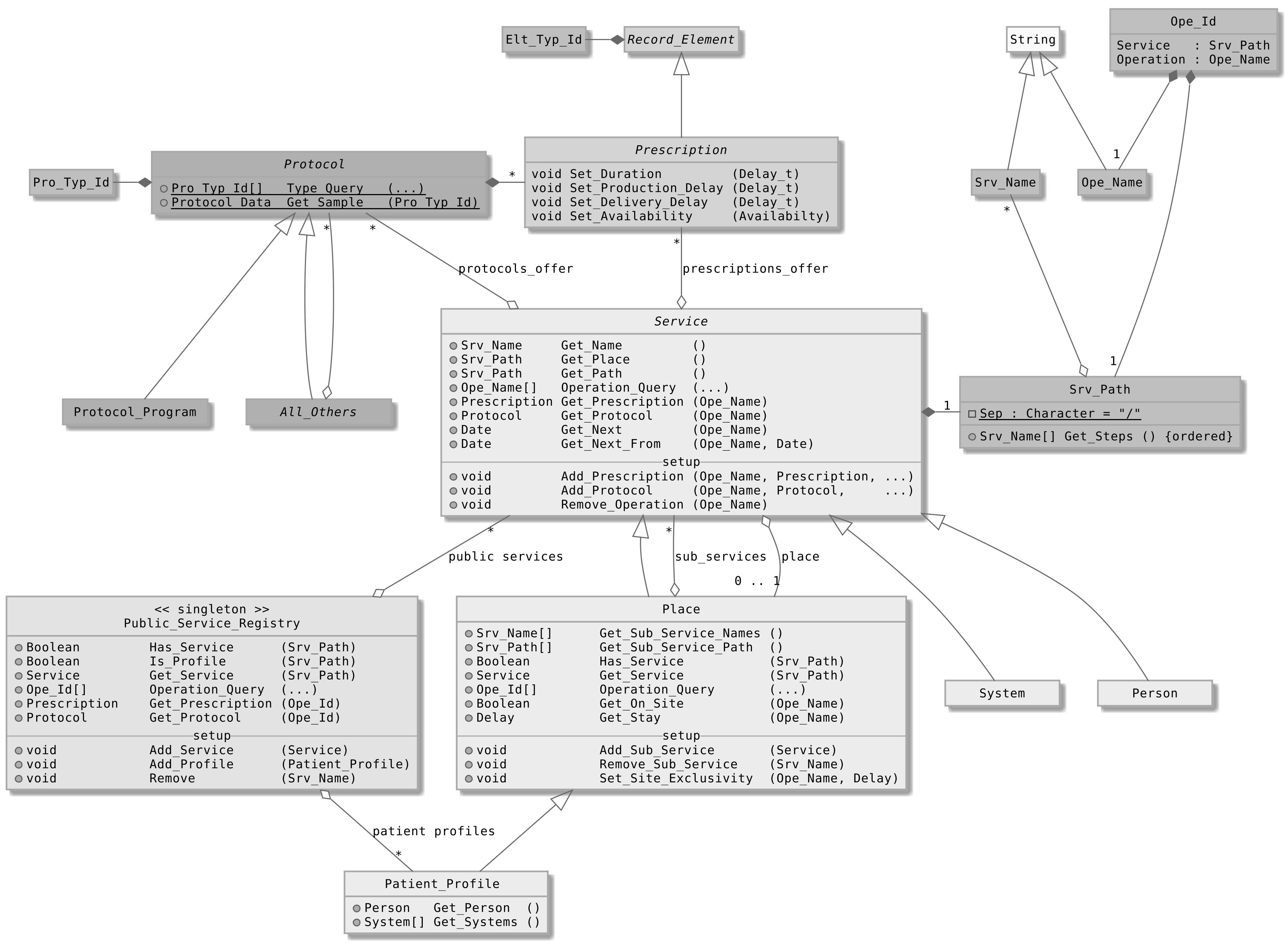}
 }
 {
 \includegraphics[width=0.9\textwidth]{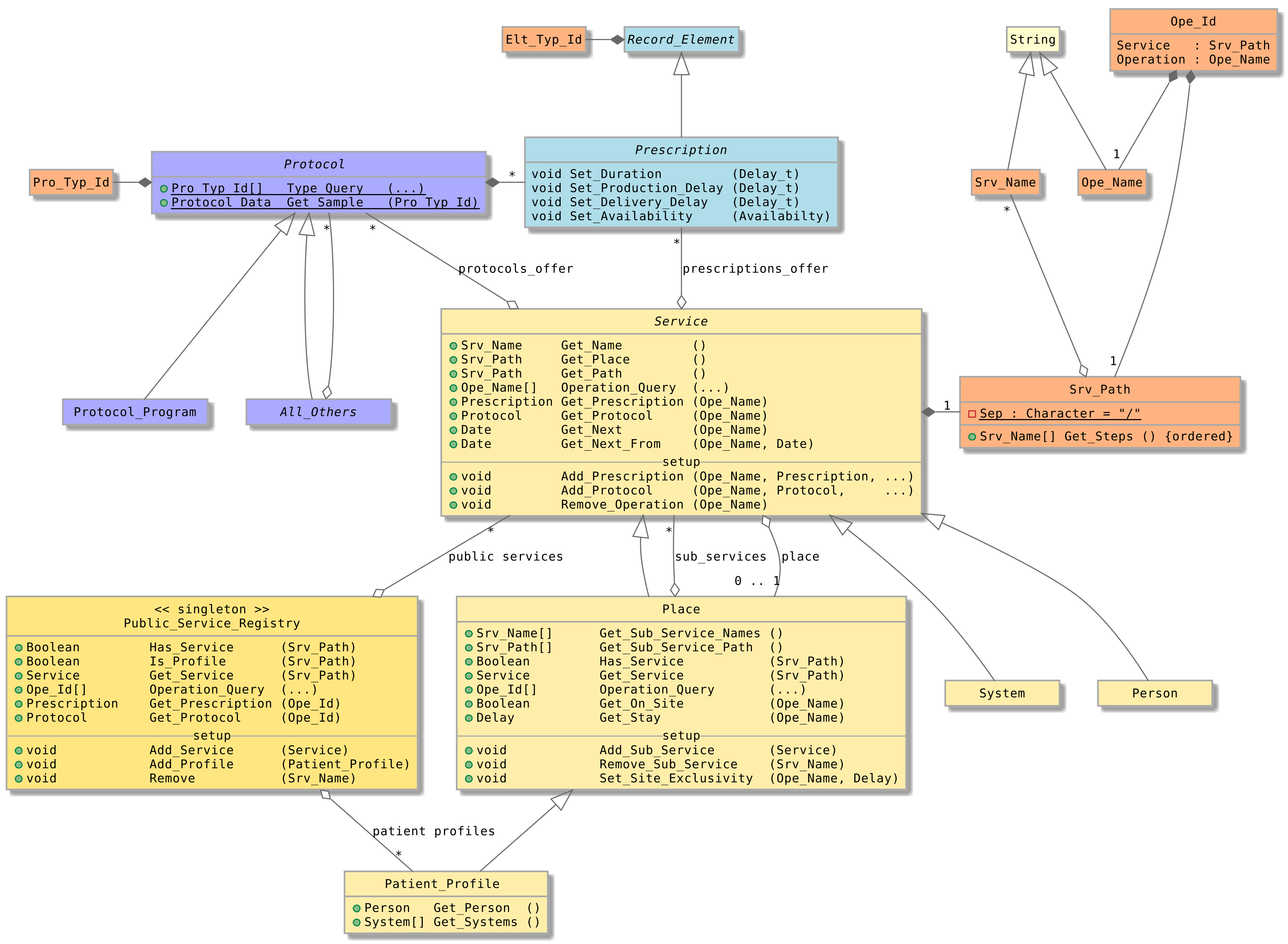}
 }
   \caption{UML class diagram of \emph{Service} showing the hierarchy of this
   class and its registration. A \emph{Service} aggregates instances of
   \emph{Protocol} and \emph{Prescription} as its offer. The class
   \emph{Service} is divided in three sub-classes as explained in section
   \ref{service-structure-and-identifier}. The class \emph{Place} is used to
   organize a service in sub-services. Identifications of service and offer
   are done locally with \texttt{Srv\_Name} and \texttt{Ope\_Name} and globally
   with \texttt{Srv\_Path} and \texttt{Ope\_Id}. The \emph{Patient\_Profile} is a
   special kind of \emph{Place} representing a patient aspect
   (\ref{patient-profile}).  It contains an unique \emph{Person} and may
   contains several \emph{Systems}.  A \emph{Service} is set up with
   \texttt{Add\_Prescription} and \texttt{Add\_Protocol} passing parametrized
   operation instances and some more parameters to define an offer contract
   (\Ref{service-contract}). The \emph{Protocol} hierarchy of this diagram
   shows the simplest one, \emph{Protocol\_Program} and the others whose
   structures can be recursive (\ref{PP}). The \emph{Public\_Service\_Registry}
   is the unique place where all public services and patient profiles are
   registered and where one can query all possibilities of offer.}
   \label{f:ServicesUML}
\end{figure*}

\begin{figure*}
\ifthenelse{\equal{\MoonArticleMode}{IEEEJBHI}}
 {
 \includegraphics[width=0.9\textwidth]{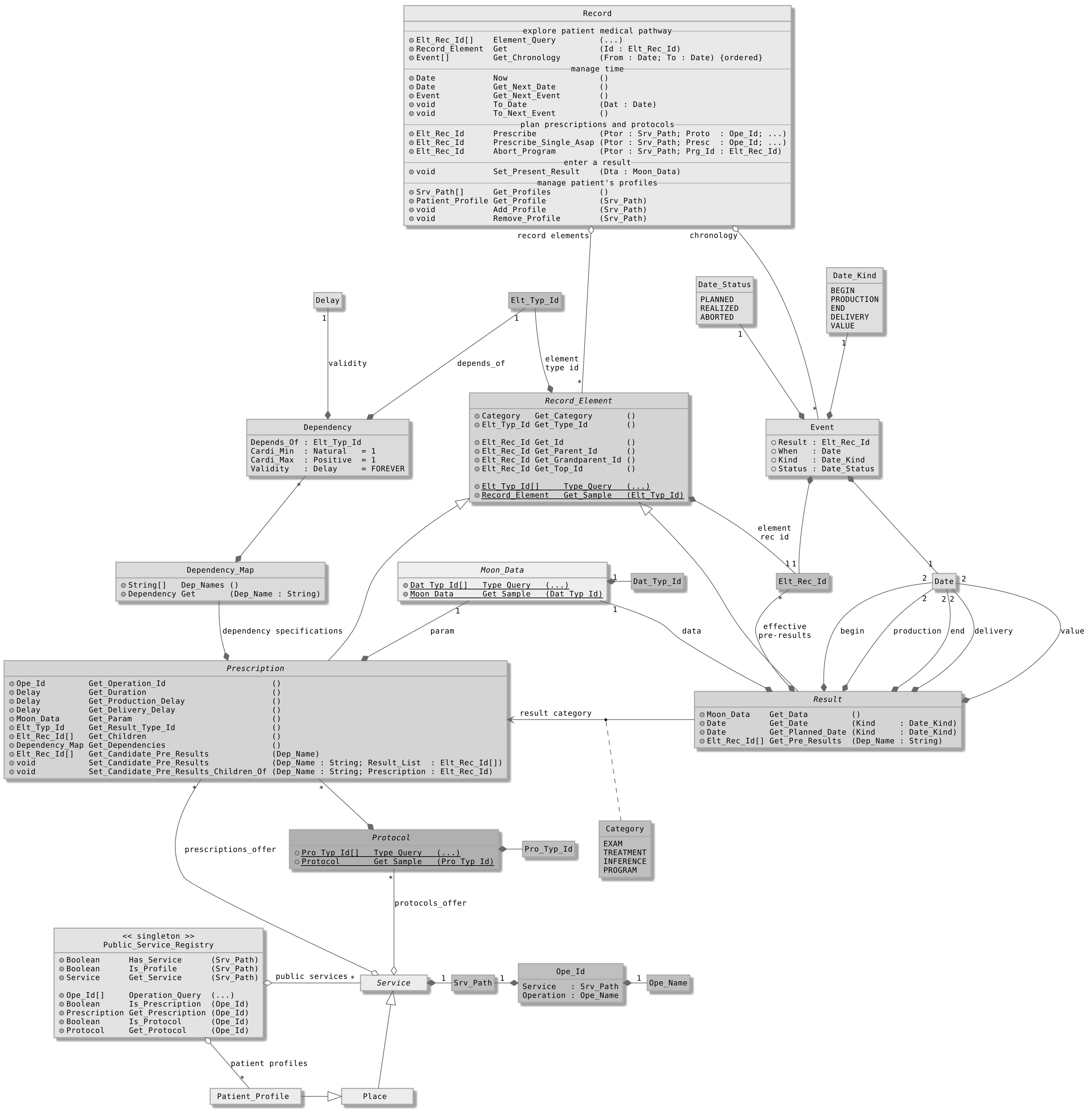}
 }
 {
 \includegraphics[width=0.9\textwidth]{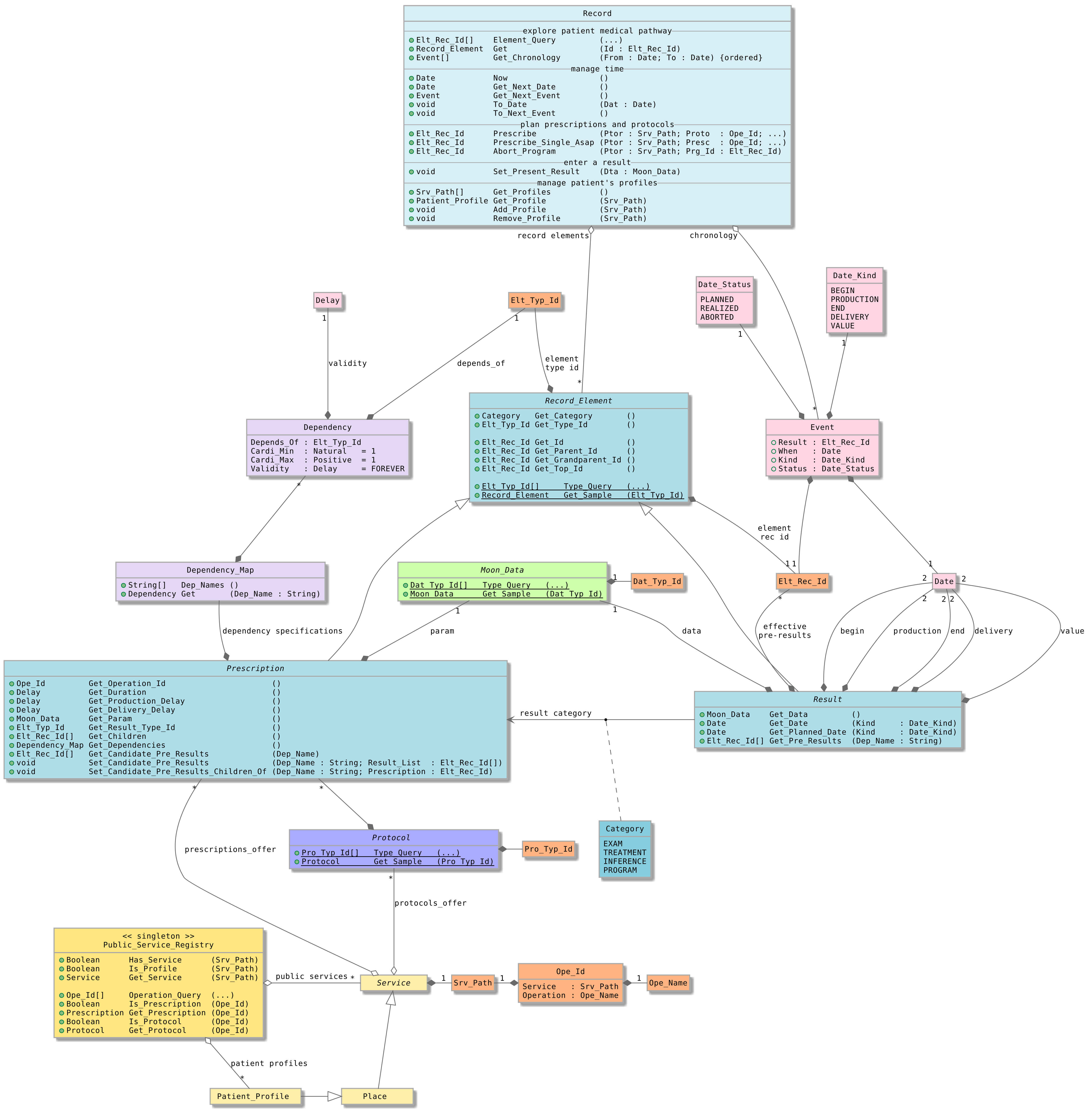}
 }
 \caption{UML class diagram of the abstract model. This diagram groups all the
   concepts of the Moon model around the \emph{Record} in a usage-oriented view.
   These usages are as follow :\\
      - \textbf{exploration of the medical model} with the static methods
      of \emph{MOON\_Data}, \emph{Record\_Element} and \emph{Protocol},\\
      - \textbf{exploration of the possibilities given by the services} using the
      \emph{Public\_Service\_Registry},\\
      - \textbf{usages of the record}, explore pathway, manage time, plan,
      fill in a result and manage profiles (\ref{record}).}
 \label{f:AbstractModelUML}
\end{figure*}

\subsection{UML Class Diagrams}
\label{class-diagrams}

In this section, the Moon abstract model is presented as UML class diagrams :

\begin{itemize}
\item
   Fig. \ref{f:MOONDataUML} shows \emph{MOON\_Data} hierarchy and specification,
\item
   Fig. \ref{f:RecElementsUML} shows \emph{Record\_Element} hierarchy and specification,
\item
   Fig. \ref{f:ServicesUML} shows \emph{Service} hierarchy and registration,
\item
   Fig. \ref{f:AbstractModelUML} shows all concepts in a usage-oriented view.
\end{itemize}

To simplify reading and highlight the organization of concepts, we have made
the following presentation choices:

\begin{itemize}
\item
   The number of class properties is reduced to a minimum to simplify reading
   and highlight organization; for example, the temporal aspects of the
   \emph{Result} class appear in the usage diagram but not in the
   specification diagram,
\item
   The pre-conditions and post-conditions of methods are not specified,
\item
   In this UML proposal, specifications are made using generic classes and thus
   directly integrate the model in the form of classes; their identifiers
   are generated by the system and are not passed as formal parameters,
\item
   The constraints associated with each of the \emph{Categories} do not appear
   in the generic specifications of the \emph{Record\_Elements},
\item
   Methods of the form \texttt{Something\_Query (...)} do not detail their
   parameters; they symbolize the many ways to perform searches,
\item
   In these diagrams, the formal parameters of the specification classes are
   reduced to a minimum; we have not included documentary details such as
   descriptions, units, keywords, etc., which are more criteria for
   distinguishing types and performing searches.
\end{itemize}


\ifthenelse{\equal{\MoonArticleMode}{IEEEJBHI}}
{\textit{Introspection of the MOON model:}}
{\paragraph{Introspection of the MOON model}}
\label{introspection}
Referencing Fig. \ref{f:AbstractModelUML}, this paragraph shows how to reach
information at different levels: medicine, services and pathway.

An actor seeking information about a symptom, pathology, exam, treatment, or
any medical practice can use the static methods \texttt{Type\_Query} of the
classes \emph{MOON\_Data}, \emph{Record\_Element}, or \emph{Protocol}. The
parameters of these methods are not specified to indicate that there are many
ways to express a query. Indeed, the introspection possibilities of these types
provide a lot of information and a multitude of combination possibilities that
can serve as search or comparison criteria.

An actor who has identified a type of prescription or protocol can, in the same
way, query the \emph{Public\_Service\_Registry} with \texttt{Operation\_Query} to
find out who offers this type of service, combining their query with criteria of
availability, location, parameterization, and practice durations. They can thus
search, compare, and select offers.
Selected offers are then scheduled using the \emph{Record}'s \texttt{Prescribe}
method for \emph{Protocols} and \texttt{Prescribe\_Single\_Asap} for
\emph{Prescriptions}.  In both cases, the prescriber provides their identifier,
the operation reference \emph{Ope\_Id}, and other informations such as patient
availability or identifiers of the preliminary results required by the offer
(\ref{scheduling-actions}).

The \emph{Record}'s \texttt{Element\_Query} method queries the record and
returns a list of identifiers of \emph{Record\_Element} instances from the
patient's pathway. The number of criteria and combinations is even greater, as
dates, dependency chains, and scheduling chains are added. The record contains
the patient's entire medical pathway, past and future, with all details. It is
possible to precisely query everything.


\ifthenelse{\equal{\MoonArticleMode}{IEEEJBHI}}
{\textit{Timing:}}
{\paragraph{Timing}}
\label{Timing}
Timing is shown at different places in the Fig. \ref{f:AbstractModelUML}.
For an instance of \emph{Prescription}, the \texttt{Get\_Duration} and
\texttt{Get\_X\_Delay} methods return its \emph{Service} setup. These delays
can be found through the dates of a planned \emph{Result} with its method
\texttt{Get\_Planned\_Date} (\ref{service-contract}). The methods
\texttt{Get\_Planned\_Date} and \texttt{Get\_Date} return respectively the
planned and the effective date of the corresponding event. When the event is
not yet passed or is aborted, these two methods return the planned date. All
these events compose a chronology stored in the \emph{Record}. Progression
within this chronology is done using the \emph{manage time} methods
(\ref{advancing-through-time}).


\ifthenelse{\equal{\MoonArticleMode}{IEEEJBHI}}
{\textit{Dependencies and Preliminary Results:}}
{\paragraph{Dependencies and Preliminary Results}}
\label{dependencies-and-preliminary-results}
In Fig. \ref{f:AbstractModelUML}, the dependencies of a \emph{Prescription}
instance are provided by the \texttt{Get\_Dependencies} method, which returns
the \emph{Dependency\_Map} resulting from the specification of the
\emph{Prescription} type (\ref{elementary-medical-specifications}).
These dependencies constrain the declaration of the prescription's preliminary
results. This declaration, necessary for its scheduling, is made for each
\emph{Dependency} with \texttt{Set\_Candidate\_Pre\_Results} to declare a list
of independent results, or with
\newline\texttt{Set\_Candidate\_Pre\_Results\_Children\_Of} to declare all the
children of a previous prescription (\ref{dependencies}, \ref{scheduling-actions}).
Once scheduling is complete, the prescription returns the entire list of
selected preliminary results with \texttt{Get\_Candidate\_Pre\_Results}.

The preliminary results of a scheduled \emph{Result} can be retrieved using its
\texttt{Get\_Pre\_Results} method. Note that the list returned by a
\emph{Result} is not the same as that returned by its \emph{Prescription}.
Indeed, the selection is made at the prescription level for all results to be
scheduled, and each \emph{Result} only select the valid subset at its
scheduling date considering the \emph{Dependencies} of its \emph{Prescription}.
So, \texttt{Get\_Pre\_Results} of a \emph{Result} is a subset of
\texttt{Get\_Candidate\_Pre\_Results} of its \emph{Presciption}.


\subsection{Medical Model}
\label{medical-model}

A medical model is a set of \emph{Record\_Element} and \emph{Protocol} classes
that specify all the activities of a medical domain
(\ref{elementary-medical-specifications}).

\begin{figure}
\begin{center}
\ifthenelse{\equal{\MoonArticleMode}{IEEEJBHI} \OR \equal{\MoonArticleMode}{IEEEJBHI_col}}
{
\ifthenelse{\equal{\MoonArticleMode}{IEEEJBHI}}{\includegraphics[width=0.98\columnwidth]{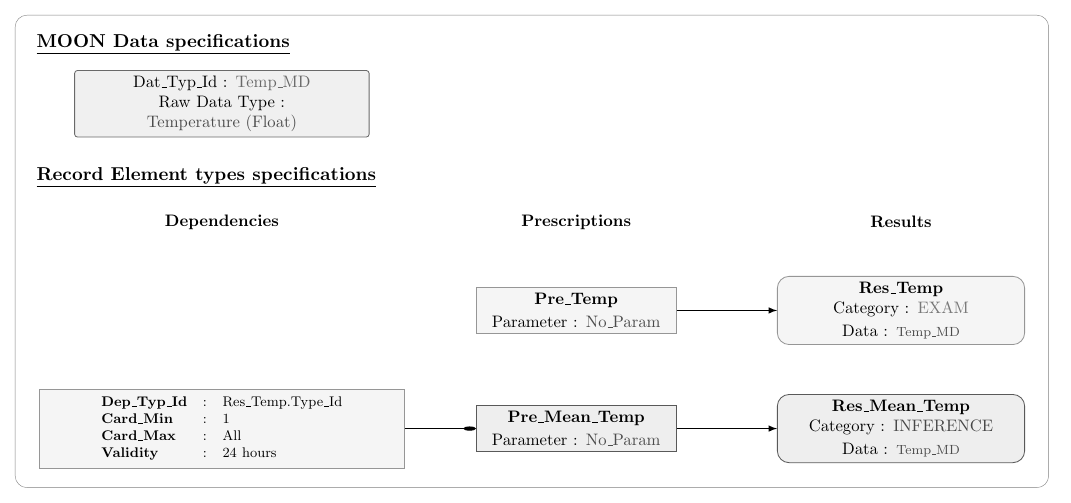}}
{\includegraphics[width=0.98\columnwidth]{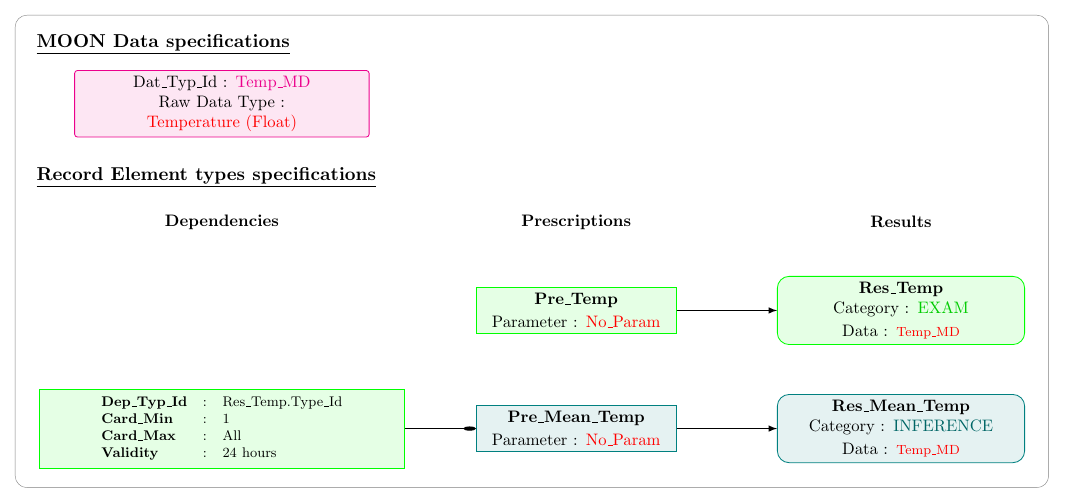}}}
{\includegraphics[width=0.98\textwidth]{tikz/temp_model_fig.pdf}}
\caption{\label{f:medical_temperature_model_scheme} Scheme of the Temperature Monitoring Model (see text for explainations).}
\end{center}
\end{figure}

Fig. \ref{f:medical_temperature_model_scheme} presents the Temperature
Monitoring Model, a toy model illustrating the specification of the activities
of a very restricted domain consisting of body temperature measurements and
the computation of their averages. This example is limited to elementary
activities since it does not include any \emph{Protocol}.  The model consists
of two \emph{Prescriptions} and two \emph{Results}, to which are added the
specifications of a \emph{Dependency} and a \emph{MOON\_Data}:

\begin{itemize}
 \item \textbf{Temp\_MD} is a body temperature value. \textbf{Temp\_MD} is
    specified as a \emph{Single} \emph{MOON\_Data} encapsulating a \emph{Float}
\item \textbf{Res\_Temp} is a temperature measurement result.
   \textbf{Res\_Temp} is specified as a \emph{Result} of category \emph{EXAM}
   carrying a \textbf{Temp\_MD},
\item \textbf{Res\_Mean\_Temp} is a result of averaging temperature
   measurements.  \textbf{Res\_Mean\_Temp} is specified as a \emph{Result} of
   category \emph{INFERENCE} carrying a \textbf{Temp\_MD},
\item \textbf{Pre\_Temp} is a temperature measurement prescription.
   \textbf{Pre\_Temp} is specified as a \emph{Prescription} without parameter
   whose result is of type \textbf{Res\_Temp},
\item \textbf{Pre\_Mean\_Temp} is a prescription for averaging temperature
   measurements over 24 hours. \textbf{Pre\_Mean\_Temp} is specified as a
   \emph{Prescription} without parameter, whose result is of type
   \textbf{Res\_Mean\_Temp}, and which has a dependency on results of type
   \textbf{Res\_Temp}, less than 24 hours old, with no cardinality limit.
\end{itemize}


\subsection{Timeline}
\label{timeline}

This section illustrates, based on the toy model, how result dates are
distributed and how the \emph{Pre\_Results} are selected.

\ifthenelse{\equal{\MoonArticleMode}{IEEEJBHI}}
{}
{\paragraph{}}
Fig. \ref{f:time_desc_exam_treat} shows an exam result \textbf{Res\_Temp} and
the dates associated with it, depending on the time settings of its
\emph{Prescription} (ref \ref{p:timing-presc-and-res}). In this case, as for
any \emph{Result} of category \emph{EXAM} or \emph{TREATMENT}, the value date
corresponds to the date the result was produced.

\begin{figure}
\begin{center}
\ifthenelse{\equal{\MoonArticleMode}{IEEEJBHI} \OR \equal{\MoonArticleMode}{IEEEJBHI_col}}
{
\ifthenelse{\equal{\MoonArticleMode}{IEEEJBHI}}
{\includegraphics[width=0.98\columnwidth]{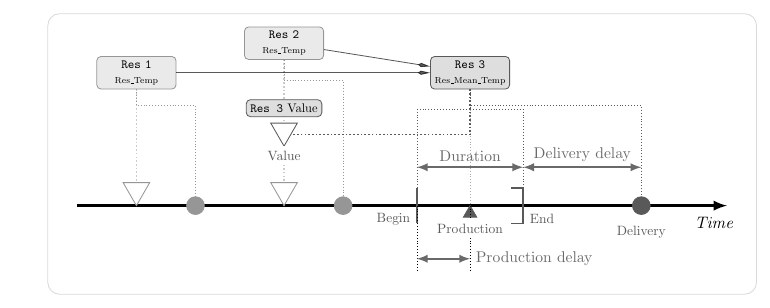}}
{\includegraphics[width=0.98\columnwidth]{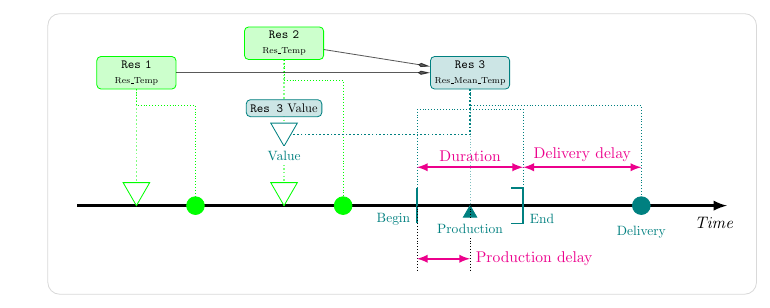}}}
{\includegraphics[width=0.98\textwidth]{tikz/inference_times_fig.pdf}}
\caption{\label{f:time_desc_infe} Description of result times for a
\textbf{Res\_Mean\_Temp} result of the Temperature Monitoring Model. The five result dates are represented in the same way as in the previous figure. The black arrows connect the inference box result to the result boxes of the \emph{Pre\_Results} of that inference. Only \texttt{Value\_Date} and \texttt{Delivery\_Date} are taken into account for the time selection of these \emph{Pre\_Results}.}
\end{center}
\end{figure}

\ifthenelse{\equal{\MoonArticleMode}{IEEEJBHI}}
{}
{\paragraph{}}
Fig. \ref{f:time_desc_infe} shows three results: two exams \textbf{Res\_Temp}
and one inference \textbf{Res\_Mean\_Temp}, which depends on the other two. These \textbf{Res\_Temp} results belong therefore to the \emph{Pre\_Results} of the inference result. In the case of
\textbf{Res\_Mean\_Temp}, as for any inference, the value date corresponds to
the value date of its last \emph{Pre\_Result}.

\ifthenelse{\equal{\MoonArticleMode}{IEEEJBHI}}
{}
{\paragraph{}}
Fig. \ref{f:Temperature_Program} illustrates the scheduling of a program
involving the two \emph{Prescriptions} of the Temperature Monitoring Model.
This program schedules a temperature measurement three times a day for 30 days,
and an average of the daily measurements at the end of each day.

The corresponding \emph{Protocol\_Program} (\ref{PP}) consists of the following
\emph{Prescription} instances: \texttt{Presc\_1} of type \textbf{Pre\_Temp} and
\texttt{Presc\_2} of type \textbf{Pre\_Mean\_Temp}, whose candidate preliminary
results are the results of \texttt{Presc\_1} (after using its
\textbf{\texttt{Set\_Candidate\_Pre\_Results\_Children\_Of}} method). The
scheduling of this protocol produces 90 instances of \textbf{Res\_Temp} and 30
instances of \textbf{Res\_Mean\_Temp}. Given this scheduling and the dependency
of \texttt{Presc\_2}, each instance of \textbf{Res\_Mean\_Temp} depends on the
three previous instances of \textbf{Res\_Temp}, and its value date is that of
the last of the three.

\newcommand{\txtResDate}[1]{\textbf{\textit{#1}}}
\begin{figure}
\begin{center}
\ifthenelse{\equal{\MoonArticleMode}{IEEEJBHI} \OR \equal{\MoonArticleMode}{IEEEJBHI_col}}
{
\ifthenelse{\equal{\MoonArticleMode}{IEEEJBHI}}
{\includegraphics[width=0.8\columnwidth]{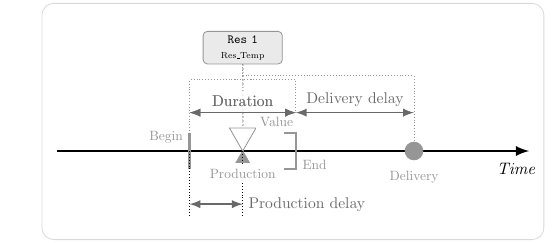}}
{\includegraphics[width=0.8\columnwidth]{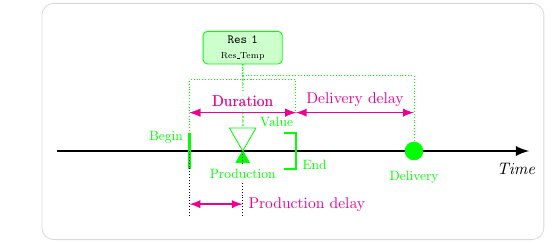}}}
{\includegraphics[width=0.8\textwidth]{tikz/exam_treat_times_fig.pdf}}
\caption{\label{f:time_desc_exam_treat} Description of
result times for a \textbf{Res\_Temp} result of the Temperature Monitoring Model. The thick
vertical line corresponds to the \txtResDate{Begin\_Date}, the thick closing bracket to the
\txtResDate{End\_Date}, the circle to the \txtResDate{Delivery\_Date}, the small filled
triangle pointing upwards to the \txtResDate{Production\_Date} and the big empty triangle
pointing downward to the \txtResDate{Value\_Date}. The \txtResDate{Duration}, the
\txtResDate{Production delay} and the \txtResDate{Delivery delay} are given by the
\emph{Service} that operates the \textbf{Pre\_Temp} prescription.}
\end{center}
\end{figure}

\begin{figure}
\begin{center}
\ifthenelse{\equal{\MoonArticleMode}{IEEEJBHI} \OR \equal{\MoonArticleMode}{IEEEJBHI_col}}
{
\ifthenelse{\equal{\MoonArticleMode}{IEEEJBHI}}
{\includegraphics[width=0.98\columnwidth]{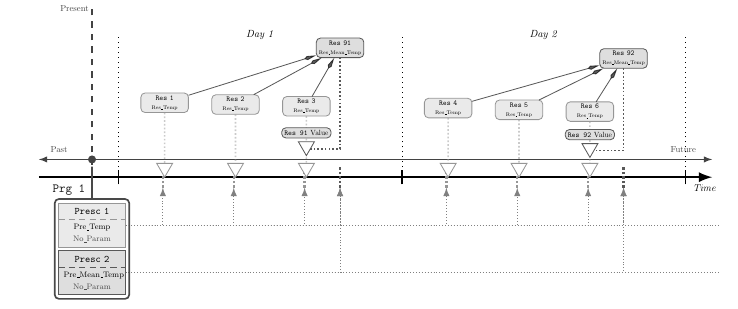}}
{\includegraphics[width=0.98\columnwidth]{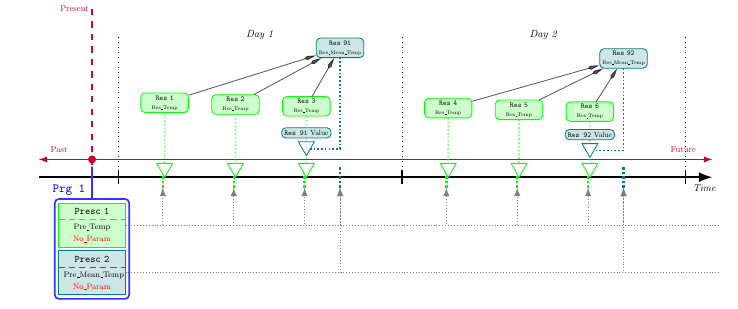}}}
{\includegraphics[width=0.98\textwidth]{tikz/program_time_fig.pdf}}
\caption{\label{f:Temperature_Program} Example of program and planned record elements. For
the sake of clarity, only the two first days are shown and only
\txtResDate{Production\_Date} and \txtResDate{Value\_Date} are displayed.
See text for further details.}
\end{center}
\end{figure}

\newpage
\section{Conclusion}
\label{conclusion}

Currently, what is called a \emph{patient's medical record} is, in fact, a set
of heterogeneous records scattered among all the stakeholders in the patient's
medical pathway. Segmenting the patient's pathway into as many records
(\emph{EMR}) as there are stakeholders poses serious organizational problems to
the detriment of the patient, who is burdened with responsibilities. These
problems generate costs, delays, errors, and ultimately health issues.

The patient's record should be shared by all stakeholders and managed
systematically to solve these management and care follow-up problems. This type
of record (\emph{EHR}) already exists and is even the subject of three
international standards.

The three EHR standards are built on a two-level architecture. The first level
consists of an abstract reference model, and the second level consists of
specific medical models proposed by the medical community. For these three
standards, the reference model provides a unified framework for record contents
but does not offer real functionalities.

Our MOON model is an abstraction of the medical pathway that can serve as a
reference model for a two-level EHR architecture. This model, in addition to
framing the data, represents the pathway across different stakeholders. It
enables numerous functionalities by itself, without necessarily resorting to
the specific models of the second level.

Modeling the pathway for any form of medicine allows for:

\begin{itemize}
\item
   Showing the entirety of past and future sequences across different stakeholders,
\item
   Rigorously managing scheduling and reservations between actors,
\item
   Managing the internal workflow of a service,
\item
   Interacting with the patient,
\item
   Performing a wide variety of simulations,
\item
   Conducting any secondary use studies of records, both prospective and retrospective,
\item
   Monitoring a population and making large-scale decisions.
\end{itemize}

Versioning of types and data, service referencing, distributed programming, and
security are not addressed by our model, but most of these aspects are already
handled by current EHR standards. We believe that
a framework like MOON, as a reference model, could enable the creation of a new standard by building
on existing EHR work, and that this standard, equipped with medical pathway
functionalities, would have a better chance of being adopted.

We have implemented MOON in a \emph{research} version oriented towards medical
simulation \cite{hommet2026}. This developing implementation has served as an
experimental basis on which we relied to structure the model and then this
article. We plan to soon present the model extensions, including
\emph{Protocols}, as well as our implementation and the simulation work carried
out.

\section*{Acknowledgment}

The authors would like to express their sincere gratitude to Phil Davis for his careful review of this article, his corrections, and his suggestions. Our fruitful exchanges have allowed us to improve this text.

--

The authors declare no conflict of interest.

--

\newpage


\bibliographystyle{unsrt}
\bibliography{moon1}

\end{document}